\documentclass[a4paper,11pt]{article}

\usepackage{jinstpub} 
\usepackage{lineno}
\usepackage{caption}
\usepackage{subcaption}
\usepackage{xcolor}
\usepackage{siunitx}
\usepackage{float}
\usepackage{ulem}
\title{\boldmath The longitudinal energy spread of ion beams extracted from an electron cyclotron resonance ion source}

\author[a,1]{J.~Angot,\note{Corresponding author.}}
\author[b,2]{O.~Tarvainen,\note{Corresponding author.}}
\author[c]{P.~Chauveau}
\author[d]{S.T.~Kosonen}
\author[d]{T.~Kalvas}
\author[a]{T.~Thuillier}
\author[a]{M.~Migliore}
\author[c]{L.~Maunoury}

\affiliation[a]{Univ. Grenoble Alpes, CNRS, Grenoble INP, LPSC-IN2P3, 38000 Grenoble, France}
\affiliation[b]{STFC ISIS Pulsed Spallation Neutron and Muon Facility, Rutherford Appleton Laboratory, Harwell, OX11~0QX, UK}
\affiliation[c]{Grand Acc{\'e}l{\'e}rateur National d'Ions Lourds (GANIL), 14076 Caen Cedex 5, France}
\affiliation[d]{Accelerator Laboratory, Department of Physics, University of Jyväskylä, Survontie 9, FI-40500 Jyväskylä, Finland}

\emailAdd{julien.angot@lpsc.in2p3.fr}
\emailAdd{olli.tarvainen@stfc.ac.uk}

\abstract{

We present a study of factors affecting the energy spread of ion beams extracted from a Charge Breeder Electron Cyclotron Resonance Ion Source (CB-ECRIS). The comprehensive simulations, supported by experiments with a Retarding Field Analyser (RFA), reveal that the longitudinal and transverse energy spread of the extracted beams are strongly affected by the electrostatic focusing effects, namely the extraction geometry and plasma beam boundary, to the extent that the electrostatic effects dominate over the magnetic field induced rotation of the beam or the effect of plasma potential and ion temperature. The dominance of the electrostatic focusing effect over the magnetic field induced rotation complicates parametric studies of the transverse emittance as a function of the magnetic field strength, and comparison of emittance values obtained with different ion sources having different extraction designs. Our results demonstrate that the full ion beam energy spread, relevant for the downstream accelerator, can be measured with the RFA only when all ions are collected. On the contrary, studying the effect of plasma properties (plasma potential and ion temperature) on the longitudinal energy spread requires heavy collimation of the beam accepting only ions near the symmetry axis of the beam for which the electrostatic and magnetic effects are suppressed. As the extraction system of the CB-ECRIS is similar to a conventional ECRIS, the conclusions of the study can be generalised to apply for all high charge state ECR ion sources. Finally, we present the results of systematic plasma potential measurements of the Phoenix-type CB-ECRIS at LPSC, varying the source potential, the microwave power and the axial magnetic field srength. It was observed that the plasma potential increases with the extraction magnetic field and the microwave power.
}


\arxivnumber{1234.56789} 

\begin{document}
\maketitle
\flushbottom

\section{Introduction}

Electron Cyclotron Resonance Ion Sources (ECRIS)~\cite{Geller_book} are used in accelerator laboratories across the globe for the production of high charge state ion beams for nuclear, atomic and material physics research and applications. Over the past decades the intensities of high charge state heavy ion beams extracted from ECRISs have increased by several orders of magnitude owing to improvements of the magnetic plasma confinement, increases in the microwave heating frequency and techniques to stabilize the plasma at high densities. The latest, superconducting, generation of these ion sources, operating at microwave frequencies $>$\SI{20}{\giga\hertz} has made beams such as uranium charge states $\geq30+$ accessible for the users (see e.g.\ ref.~\cite{Leitner08}). The versatility and continuous improvement of ECR ion sources have cemented their role as the workhorse particle injectors into cyclotrons and linac accelerators as well as charge breeders (CB) at radioactive beam facilities.

The main parameters of the ion beams extracted from ECR ion sources (or any other ion source) are the ion species (mass and charge state), the beam energy and energy spread, the intensity (current), and the beam quality, which is often quantified using the beam brightness,
\begin{equation}
b=\frac{I}{\epsilon_x\epsilon_y},
\end{equation}
where $I$ is the beam current, and $\epsilon_x$ and $\epsilon_y$ are the horizontal and vertical transverse (to beam propagation) emittances. It is also common to characterise the beam quality with the two transverse emittances, $\epsilon_x$ and $\epsilon_y$, owing to the fact that only the fraction of the beam inside the transverse acceptance of the accelerator is of any use. The transverse emittance of the beam is a measure of the relative position and divergence (e.g.\ $x$ and $x'=\frac{p_x}{p_z}$ or $y$ and $y'=\frac{p_y}{p_z}$) of the beam particles.

The transverse emittances are used for describing the beam quality in the ECRIS context almost exclusively, albeit they correspond to only two of the 2D projections, namely $f_x(x,x')$ and $f_y(y,y')$, of the corresponding position and momentum distribution function $f(x,y,z,p_x,p_y,p_z)$ of the complete 6D phase space representation of the ion beam. Typically the transverse phase space distributions are measured with devices such as Allison scanners~\cite{Allison83}, while measurement of 4D $f_{xy}(x,y,p_x,p_y)$~\cite{Kremers2013} and even full 6D phase spaces have been demonstrated~\cite{Cathey18}. The 2D projection, $f_z(z,p_z)$, describes the distribution of particle longitudinal momenta and locations, which in the case of continuous ion beams reduces to a distribution of the longitudinal momentum, presented as kinetic energy in this paper. The longitudinal energy spread of the beam originates from two sources, firstly from the variation of the particle total kinetic energy (variation of source potential) and secondly from the longitudinal momentum converted to transverse momentum due to focusing effects conserving the total momentum, i.e. $p_z^2=p_\text{tot}^2 - p_x^2 - p_y^2$. The latter effect typically varies along the beam transport, while the former is constant and affects the matching of the ion beam into the subsequent accelerator because it couples into the transverse momentum spread in dispersive ion optical components, e.g.\ bending magnets required for the $m/q$-analysis of the (total) ion beam extracted from the ECRIS, thus increasing the transverse emittance in the bending plane of the magnet.

\begin{figure}[!htb]
    \centering
    \includegraphics[width=0.95\textwidth]{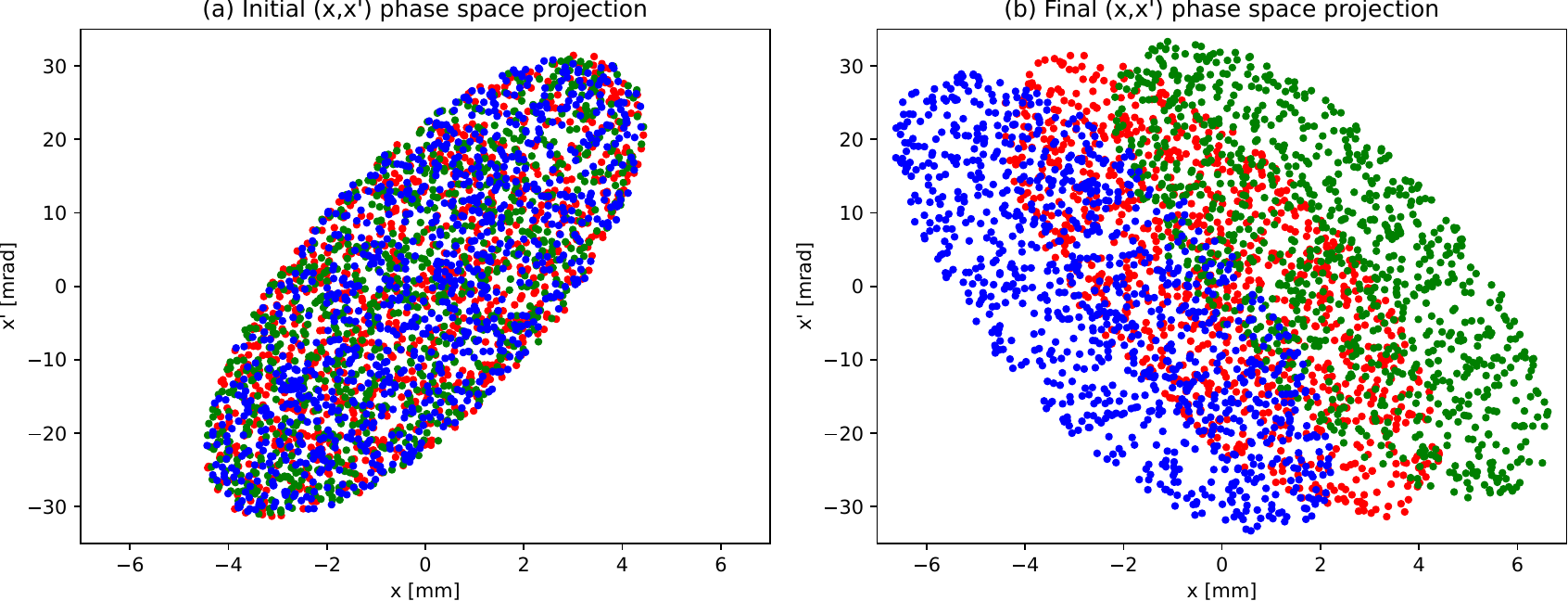}
    \caption{The effect of a bending magnet on the ($x,x'$) phase space projection of an ion beam with non-zero kinetic energy spread. (a) Initially the particles with varying kinetic energies (red dots have 20~keV, blue 19.94~keV and green 20.06 keV) are distributed uniformly in the phase space. (b) After the dispersive beamline component the beam occupies a larger total area in the ($x,x'$) phase space in comparison to those ions with kinetic energy of $20$ keV, i.e.\ the corresponding transverse emittance has increased due to the kinetic energy spread.}
    \label{emittance_schematic}
\end{figure}

The effect of the two longitudinal energy spread components is best demonstrated by calculating the propagation of an ion beam through a typical 90$^\circ$ double-focusing bending magnet. Figure~\ref{emittance_schematic} presents the initial and final horizontal $(x,x')$ phase space distributions of the beam with three components. All particles are $^{16}$O$^{+}$. The beam components have total kinetic energies of 20~keV, 19.94~keV and 20.06~keV, i.e.\ the energy spread is $\pm 60$~eV -- an arbitrary value chosen to demonstrate the effect. The beam has an initial transverse emittance of $100 \pi$~mm~mrad and is initially diverging, introducing a further longitudinal energy spread. After the bending magnet, the beam is converging due to the focusing force of the magnet and the dispersion of the magnet has caused a separation of the different kinetic energy components. The transverse emittance of the beam has increased, while the 6D emittance of the beam is conserved according to Liouville's theorem as the longitudinal momentum spread is converted to transverse momentum spread. Typically the effect on the transverse beam emittance is not significant, but nevertheless, the effect can be mitigated by designing an achromatic low energy beam transport line. Such systems have been used to accelerate two neighbouring charge states of heavy ions in modern accelerators \cite{Ostroumov2008}.

As demonstrated here, the kinetic energy spread of the ion beam causes the beam to occupy a larger transverse phase space area, which affects its transport and acceleration. This serves as a motivation to identify various mechanisms influencing the longitudinal energy spread of an ECRIS and to refine techniques for the measurement of the energy spread, which is the topic of this paper. 

\section{Transverse and longitudinal energy spread of ion beams extracted from an ECRIS} \label{theory}

Positive ion beams are extracted from an ECRIS by introducing an electric field between the plasma electrode (at ion source potential) and a puller electrode (typically at ground potential or at negative bias). The outstanding feature of an ECRIS is that the ions are extracted from a strong axial magnetic field with a sextupole component, which affects the division of longitudinal and transverse energies downstream from the ion source at zero magnetic field, resulting in relatively large $\Delta p/p$. Here, we identify different mechanisms affecting the longitudinal energy spread of the ion beams extracted from an ECRIS. The importance of each mechanism is then studied with \textsc{Simion}~\cite{Simion} and \textsc{IBSimu}~\cite{Kalvas_IBSimu} simulations described in the following sections.

The kinetic energy $K$ of the ions extracted from an ECRIS is defined by the ion charge $q$, applied ion source potential $V_s$ and the induced ambipolar plasma potential $V_p$, i.e.\
\begin{equation}
    K=q(V_s+V_p).
    \label{potential}
\end{equation}

The ambipolar potential, which is typically 5--\SI{50}{\volt} builds up to balance the mobilities and loss rates of positive and negative charges from plasma, thus maintaining its quasi-neutrality~\cite{Riemann91}. Not all ions originate from the maximum plasma potential, but instead the plasma potential has a spatial profile and ions traversing the plasma pre-sheath can experience collisions, which together results in an inherent longitudinal energy spread $\Delta K_l$ such that:

\begin{equation}
    \frac{\Delta K_l}{K}=\frac{V_p}{V_s}.
\end{equation}

The energy spread corresponding to the ambipolar plasma potential is found at energies $K>qV_s$ i.e. the minimum total energy of the ions is defined by the source potential.

\begin{figure}[hbt]
     \centering
     \begin{subfigure}[b]{0.48\textwidth}
         \centering
         \includegraphics[width=\textwidth]{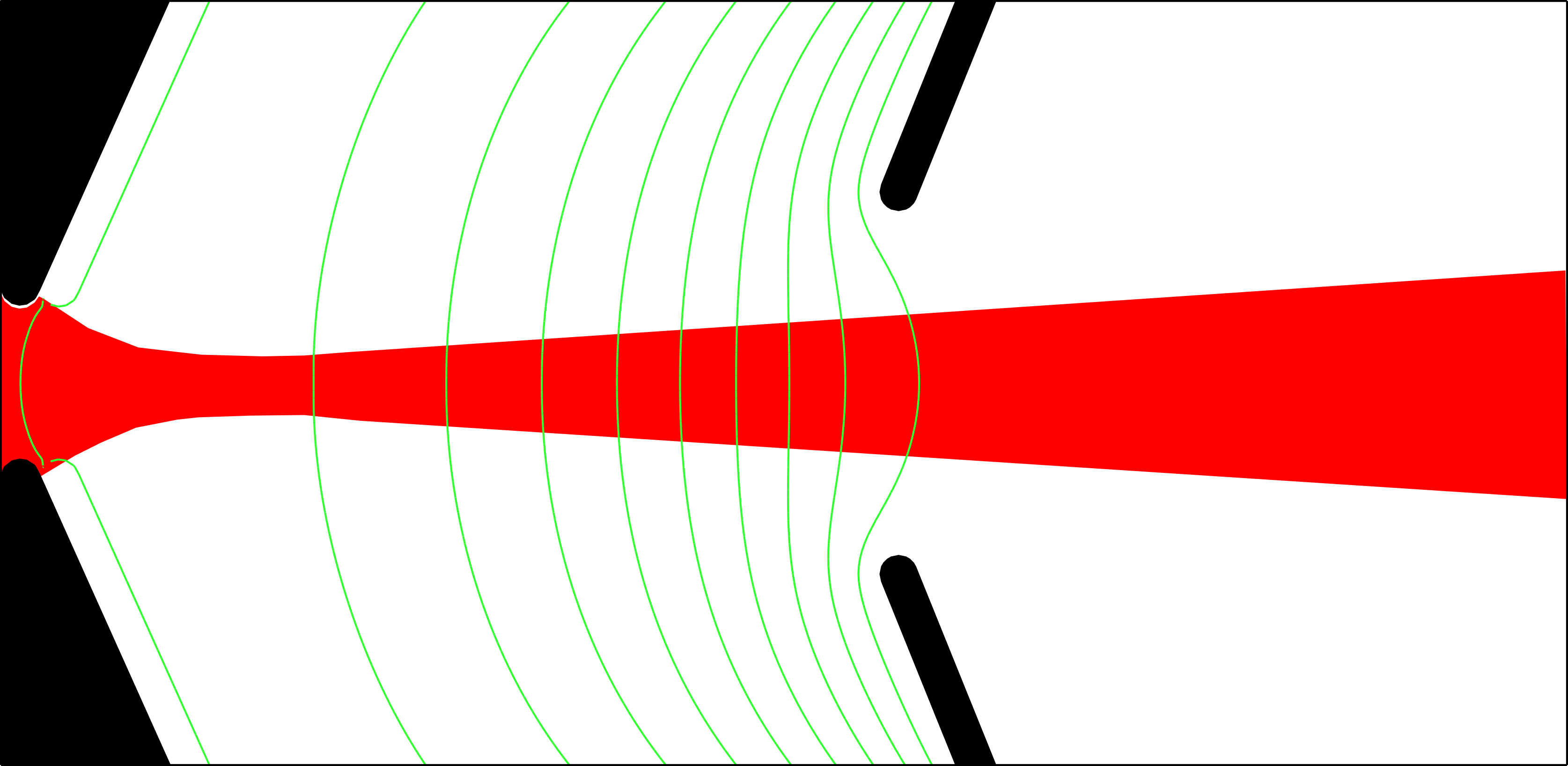}
         \caption{Real extraction}
     \end{subfigure}
    \hfill
     \begin{subfigure}[b]{0.48\textwidth}
         \centering
         \includegraphics[width=\textwidth]{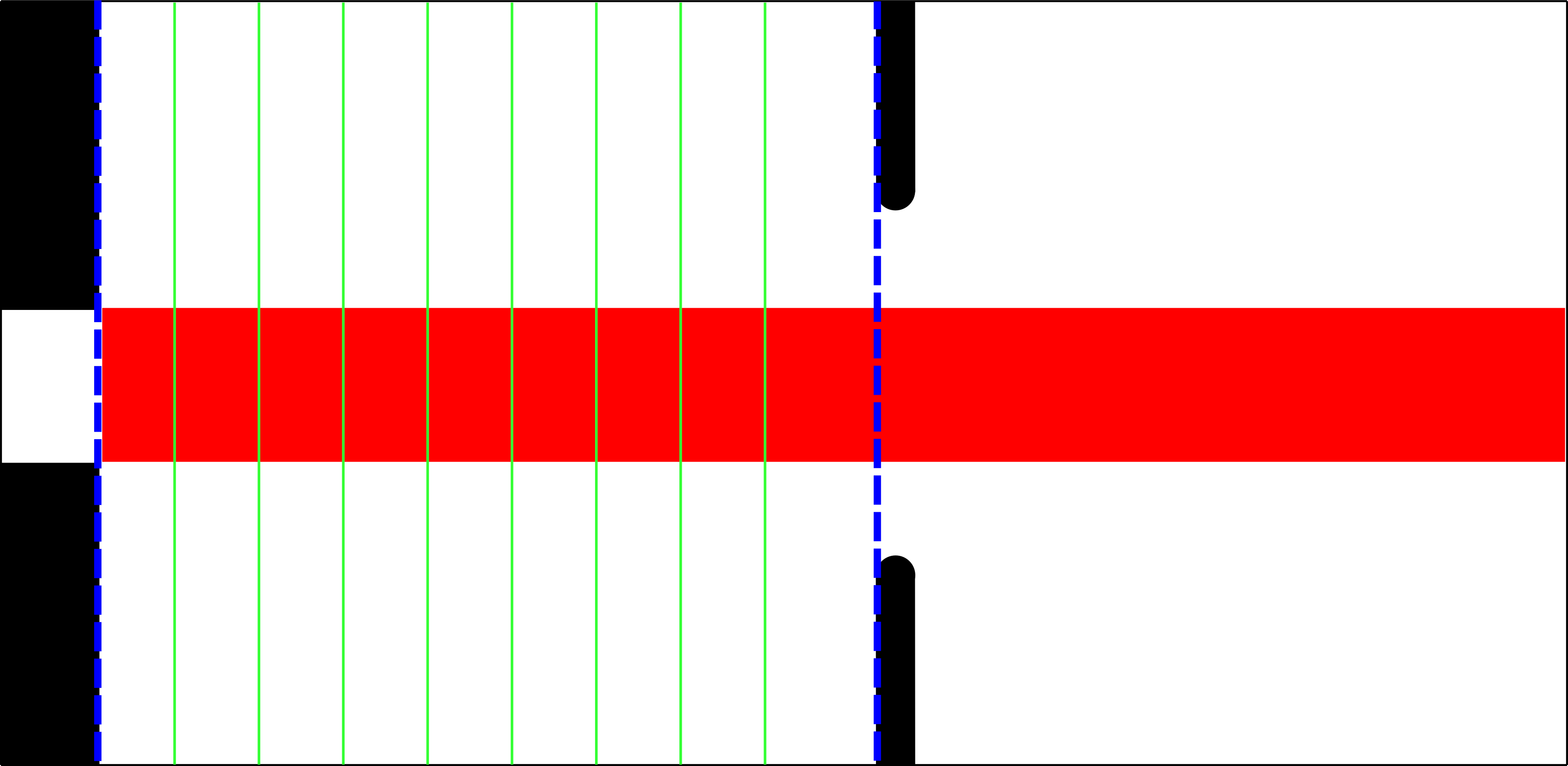}
         \caption{Parallel extraction}
     \end{subfigure}
    \caption{Schematic figure of the LPSC CB extraction system (a) illustrating the curvature of the equipotential surfaces and (b) a modified extraction with the same distance between the electrodes and artificially created parallel equipotential lines (the dashed blue lines represent the modified geometry used to create the parallel system).}
    \label{fig:parallel_vs_real_extraction}
\end{figure}

In ideal extraction systems the electric field propelling the ions to the accelerator's injection energy is parallel to the beam axis ($E=E_z$). This is rarely the case in practical extraction geometries where the electric field often has a radial component due to the curvature of the plasma-beam boundary (often referred to as the plasma meniscus) and the electrodes shaping the electric field. As an example, figure~\ref{fig:parallel_vs_real_extraction}(a) shows the extracted ion beam and the projection of the equipotential surfaces in the LPSC Phoenix CB extraction system, used in the simulations in the following section where it will be described in detail. The real extraction geometry is compared to an ideal system, where the electric field has only an $E_z$-component, shown in figure~\ref{fig:parallel_vs_real_extraction}(b).

The concave shape of the equipotential surfaces implies a radial electric field component $E_r$, focusing the beam and resulting in transverse kinetic energy $K_t$. Since the total ion energy is defined by the source (+ plasma) potential, the transverse energy reduces the longitudinal energy of the beam. Furthermore, as the curvature of the equipotential surfaces depends on the radial coordinate, the electrostatic focusing effect results in both, longitudinal and transverse, energy spread. While the ambipolar plasma potential results in kinetic energies $K>qV_s$, the energy spread caused by the electrostatic focusing effect is observed at longitudinal energies $K_l<q(V_s+V_p)$.

Another process causing longitudinal energy spread at energies $K_l<qV_s$ is the azimuthal (transverse) thrust experienced by the particles propagating towards slowly decreasing solenoid magnetic field~\cite{Brewer}, which creates a beam rotation. This phenomenon is theoretically described by the so-called Bush theorem and is usually referred to as the magnetic emittance. In modern ECR ion sources the magnetic field typically fulfills the semi-empirical magnetic scaling laws~\cite{Hitz01}, which implies that the extraction field is approximately 1.8 times the resonance field, i.e. the field strength scales with the microwave frequency used for the plasma (electron) heating. The resulting radial velocity $v_r$ of the beam at zero magnetic field far away from the ion source can be expressed as
\begin{equation}
    v_r=-\frac{r_0q^2}{4m^2v_z}\int B^2dz=\frac{qBr_0}{2m},
\end{equation}
where $B$ is the magnetic field at the extraction and $r_0$ the initial radial coordinate of the particle with charge $q$ and mass $m$. As the energy of the particles is conserved in the magnetic field decreasing along the beam path, the emergence of the transverse velocity component reduces the longitudinal energy of the ions by
\begin{equation}
   \frac{\Delta K_l}{K}=\frac{qB^2r_0^2}{8m(V_s+V_p)+qB^2r_0^2},
   \label{rotation}
\end{equation}
which results in corresponding longitudinal (and transverse) energy spread. The above expression does not take into account the transverse velocity of the ions due to the electrostatic focusing in the extraction gap. Similar to the electrostatic focusing effect, the longitudinal energy spread caused by the extraction from a solenoid field, is found at energies $K_l<q(V_s+V_p)$, i.e.\ the two effects are superimposed. Their relative importance depends on the extraction geometry, the effective extraction radius of the ions and the strength of the magnetic field at the extraction.

Thus far we have ignored the fact that the ions are not stationary when they reach the plasma-beam boundary. In reality they have a distribution of transverse velocities, i.e.\ ion temperature, which according to spectroscopic measurements can be more than \SI{10}{\electronvolt} in ECRIS plasmas~\cite{Kronholm19}. The associated transverse energy can be converted to longitudinal energy as the extracted ions propagate towards decreasing solenoid field. This follows from the quasi-conservation of the magnetic moment, 
\begin{equation}
    \mu=\frac{mv_\perp^2}{2B}=\frac{K_\perp}{B}\approx\textrm{constant},
    \label{moment}
\end{equation}
implying that the transverse velocity (energy) of the extracted ions decreases with the magnetic field until all of the initial transverse energy is converted to longitudinal energy. The contribution of the initial transverse velocity (ion temperature) on the final longitudinal energy spread does not depend on the initial radial coordinate of the particle. Thus, the contribution of the ion temperature on the longitudinal energy spread is larger than the contribution of the azimuthal thrust or the electrostatic focusing for ions extracted near the symmetry axis, i.e. when $r_0 \ll R$, where $R$ is the radius of the plasma electrode aperture. In other words, due to their initial temperature, the ions extracted near the axis can be expected to have a spread of longitudinal energies at $K_l>q(V_s+V_p)$.  

The above fundamental processes affecting the longitudinal energy spread of the ion beams extracted from an ECRIS are accompanied by effects such as space charge forces, plasma-beam boundary (meniscus), the sextupole field component in the extraction and the variation of the initial particle distribution (effective beam radius at the outlet aperture), which are difficult if not impossible to quantify analytically. Each of these effects are addressed hereafter through simulations.

As stated above (see eq.~\ref{potential}), the ion beam extraction requires the polarisation of the source at high-voltage ($V_s$), typically several tens of \SI{}{\kilo\volt}. This is achieved with high-voltage power supplies with a temporal stability on the order of 0.01-0.1\% (given by the manufacturer). The output voltage ripple $\rho$ is typically at high frequency, on the order of several tens of \SI{}{\kilo\hertz}. This induces a time-averaged energy spread of the extracted beam $\Delta K=\rho q V_s$. As an example, a 0.05\% ripple at \SI{20}{\kilo\volt} translates to an energy spread of $q \times$\SI{10}{\electronvolt}. This phenomenon is related to technical limitations and will be subsequently compared to the energy spread caused by the aforementioned physical processes.

\section{Extraction simulations}

In order to assess the relative importance of the different processes affecting the longitudinal energy spread of ion beams extracted from ECR ion sources we conducted an extensive simulation study. The majority of the simulations were carried out with \textsc{Simion} software~\cite{Simion} with some of the cases studied with \textsc{IBSimu}~\cite{Kalvas_IBSimu} to account for possible plasma extraction and space charge effects.

\subsection{\textsc{Simion} extraction simulations}
\label{SimionSimu}

\subsubsection{Model}
\begin{figure}[htp]
    \centering
    \includegraphics[width=\textwidth]{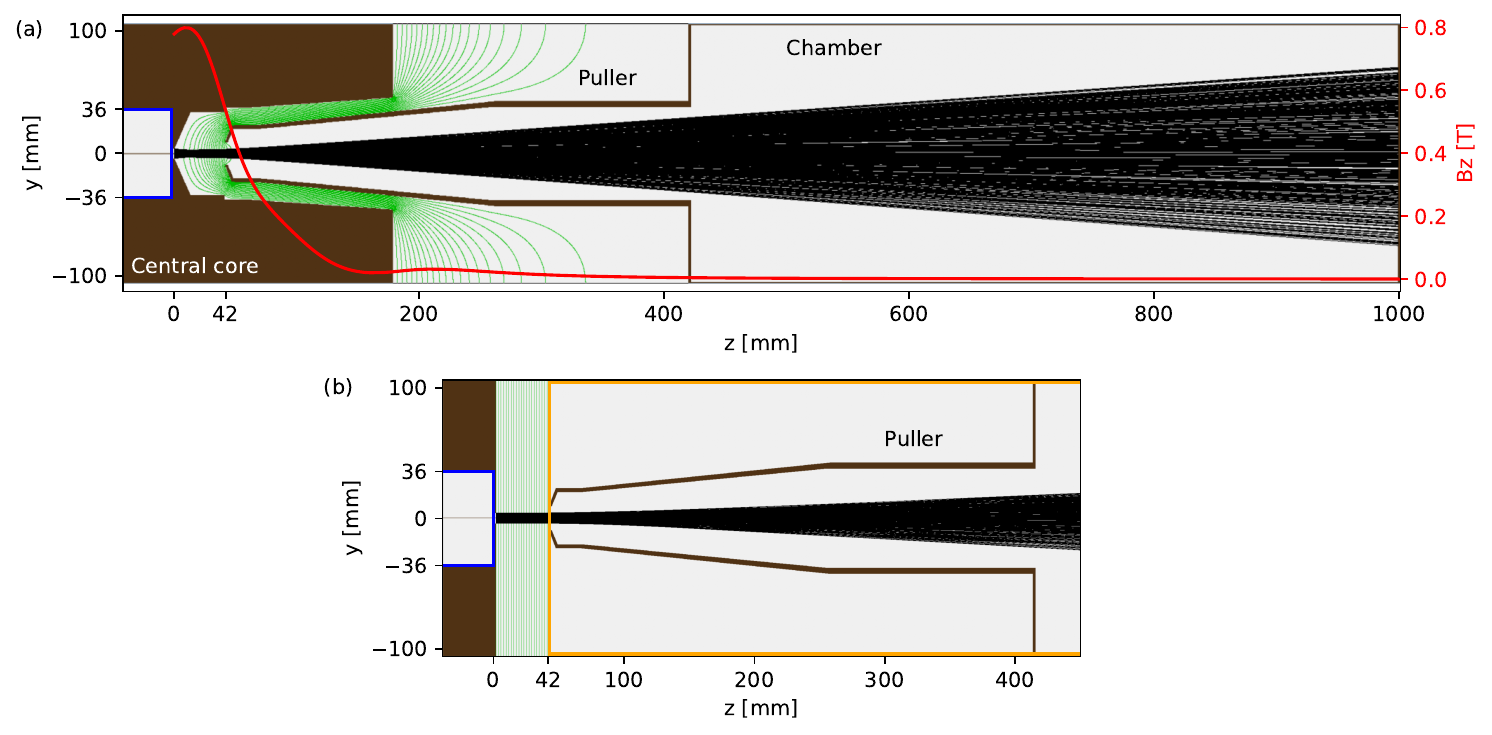}
    \caption{(a) The \textsc{Simion} model of the LPSC Phoenix CB ECRIS extraction with the (artificial) \SI{20}{\kilo\volt} electrode in blue, equipotential lines in green, ion trajectories in black, solenoidal axial magnetic field on axis superimposed in red and (b) partial view of the modified model with the artificial ground electrode in orange.}
    \label{fig:extraction models}
\end{figure}

\textsc{Simion} was used for the majority of simulations without considering space charge effects. The goal was twofold; to determine the parameters that have the most significant influence on the longitudinal energy spread of the extracted ions, and a quantitative comparison to the above, purely theoretical, estimates of the energy spread. A model of the LPSC Phoenix CB extraction shown in figure~\ref{fig:extraction models} (a) was used. The model accounts for the electrode geometry and the resulting electrostatic potential distribution.

Similar to the experiments described later, a \SI{20}{\kilo\volt} potential was applied to the plasma chamber and plasma electrode whereas the puller electrode and the extraction chamber were set to ground potential. The round aperture diameters are \SI{8}{\milli\meter} plasma electrode and \SI{18}{\milli\meter} puller, the (extraction) gap between the two is \SI{42}{\milli\meter}. Figure~\ref{fig:extraction models} (a) shows the electrostatic equipotential lines in green, corresponding to values from 2 to \SI{20}{\kilo\volt}, with \SI{2}{\kilo\volt} step. A flat emitting surface was considered in the model. This was achieved by generating a constant potential of \SI{20}{\kilo\volt} at the plasma electrode aperture with an artificial \SI{20}{\kilo\volt} infinitely thin electrode located inside the plasma chamber. As such the model does not account for the dynamic nature of the true plasma-beam boundary, which was addressed with the \textsc{IBSimu} simulations described later. 

In both, the \textsc{Simion} and \textsc{IBSimu} models, the magnetic field was taken into account by superimposing 3D field maps (with varying field strengths), previously calculated using \textsc{Radia3D} software~\cite{Radia3D}. The magnetic structure consists of (i) the three coils and return yoke, (ii) the injection and extraction magnetic steel plugs assembled into the central core, and (iii) the permanent magnet sextupole creating \SI{0.8}{\tesla} radial field at the plasma chamber wall in front of the magnet poles \cite{Angot2020}. The field maps were calculated with coil currents typically used for charge breeding operation. Three different current values were applied to the extraction coil to obtain an axial extraction field on axis of \SI{0.5}{\tesla}, \SI{0.8}{\tesla} and \SI{1.1}{\tesla}, keeping the other parameters unchanged. This was done to study the effect of the extraction field strength on the longitudinal energy spread at the end of the simulation domain. For these simulations, a long extension of the magnetic map, up to \SI{1000}{\milli\meter} from the plasma electrode towards the beam propagation direction ($z$) was created. At this distance the magnetic field is negligible ($<4\cdot 10^{-4}$~T in the \SI{0.8}{\tesla} 3D map case) with respect to the field at the plasma electrode. The dimension of the 3D map was then \SI{212}{\milli\meter} $\times$ \SI{212}{\milli\meter}$\times$ \SI{1008}{\milli\meter}, with a \SI{2}{\milli\meter} Cartesian grid.

For each simulation, 10000 ions were launched from a disc concentric with the plasma electrode aperture, towards the puller ($z$-coordinate). The varied parameters were the ion charge state, ion mass, initial longitudinal energy (representing the plasma potential), initial transverse energy spread (representing the ion temperature), and the emitting disc diameter. The disc diameter was varied to change the effective extraction radius, which is known to depend on the charge state of the ions~\cite{Wutte, Panitzsch}. The ion velocity components were recorded at the end of the simulation domain ($z=$\SI{1000}{\milli\meter}) and their longitudinal and transverse energies were calculated from these values.

In order to separate the effects of the electrostatic focusing and the magnetic field induced rotation on the ion energy spread, another \textsc{Simion} model was implemented to obtain parallel equipotential surfaces between the plasma electrode and the puller. Such geometry minimises the contribution of electrostatic focusing but maintains the magnetic field effects. This was achieved by modifying the geometry to create a flat surface at the plasma electrode plane and adding an artificial ground electrode starting at the plane where the tip of the puller would normally locate. The simplified extraction geometry is shown in figure~\ref{fig:extraction models} (b). 

\subsubsection{Results}
In the first simulations $^{16}$O$^+$ ions were launched from a \SI{8}{\milli\meter} diameter disc (plasma electrode aperture) with \SI{20}{\kilo\volt} source potential and \SI{0.8}{\tesla} solenoidal magnetic field (maximum) at the extraction. The ions were launched without any initial energy, i.e.\ the plasma potential and ion temperature were omitted. This parameter set, listed in table~\ref{tab:reference}, was considered as the reference point for subsequent simulations probing the effects of each parameter on the longitudinal energy spread of the ion beam, i.e. the simulation parameters are those listed here unless otherwise stated.

\begin{table}[H]
    \caption{Reference parameters of the extraction simulations.}
    \centering
    \begin{tabular}{|c|c|}
    \hline
    Ion mass (amu)            & 16 \\
    Ion charge (e)          & 1 \\
    Emitting disc diameter (\SI{}{\milli\meter}) & 8 \\
    B-field at extraction (\SI{}{\tesla})  & 0.8 \\
    Ion source potential (\SI{}{\kilo\volt})  & 20 \\
    Initial longitudinal energy (\SI{}{\electronvolt})     & 0 \\
    Ion temperature (\SI{}{\electronvolt})    & 0 \\
    \hline
    \end{tabular}
    \label{tab:reference}
\end{table}

\paragraph{Magnetic field}

\begin{figure}[htp]
    \centering
    \includegraphics[width=\textwidth]{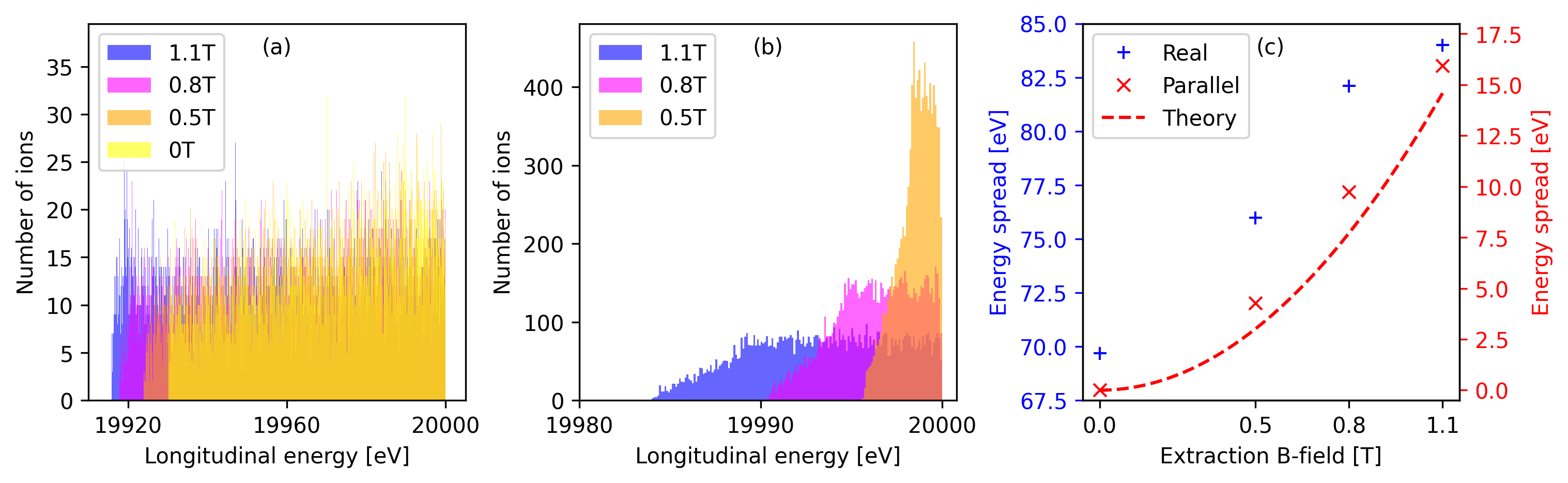}
    \caption{The longitudinal energy distribution of $^{16}$O$^+$ ion beam as a function of the extraction magnetic field strength for (a) the LPSC CB geometry (real), and (b) simplified extraction geometry without electrostatic focusing. The maximum energy spread of the ions in these cases is shown in (c).}
    \label{fig:real vs parallel}
\end{figure}

Figure~\ref{fig:real vs parallel} compares the longitudinal energy distributions of $^{16}$O$^+$ ions found at different values of the extraction magnetic field (maximum axial field), in the cases of the real extraction (a) and parallel electric field line model (b). In the case where no magnetic field was applied for the parallel model, all ions exit the simulation domain with a \SI{20}{\kilo\electronvolt} longitudinal energy. The maximum energy spread of the ions in these cases are shown in figure~\ref{fig:real vs parallel}(c) together with the theoretical estimate from eq.~\ref{rotation} for the parallel E-field case.

The comparison of the results (figure~\ref{fig:real vs parallel}) obtained with the two extraction geometries reveals that in the case of the LPSC CB, the longitudinal energy spread of the ions is dominated by the electrostatic focusing in the extraction system. The total longitudinal energy spread of the beam is approximately \SI{70}{\electronvolt} even without applying the magnetic field, increasing to approximately \SI{85}{\electronvolt} at \SI{1.1}{\tesla} extraction field. Hence, the maximum contribution of the magnetic field is approximately 20\% of the electrostatic focusing effect. The effects of the electrostatic focusing and the magnetic field strength both contribute to the final energy spread, which is a superposition of the two effects. The effect of the magnetic field is best visible with the idealised parallel E-field extraction, where it results in 4--15~eV energy spread with 0.5--1.1~T extraction field.  The longitudinal energy spread caused by the rotation of the beam in the magnetic field is proportional to the field strength as predicted by the theory. 

\paragraph{Ion mass}

\begin{figure}[htp]
    \centering
    \includegraphics[width=\textwidth]{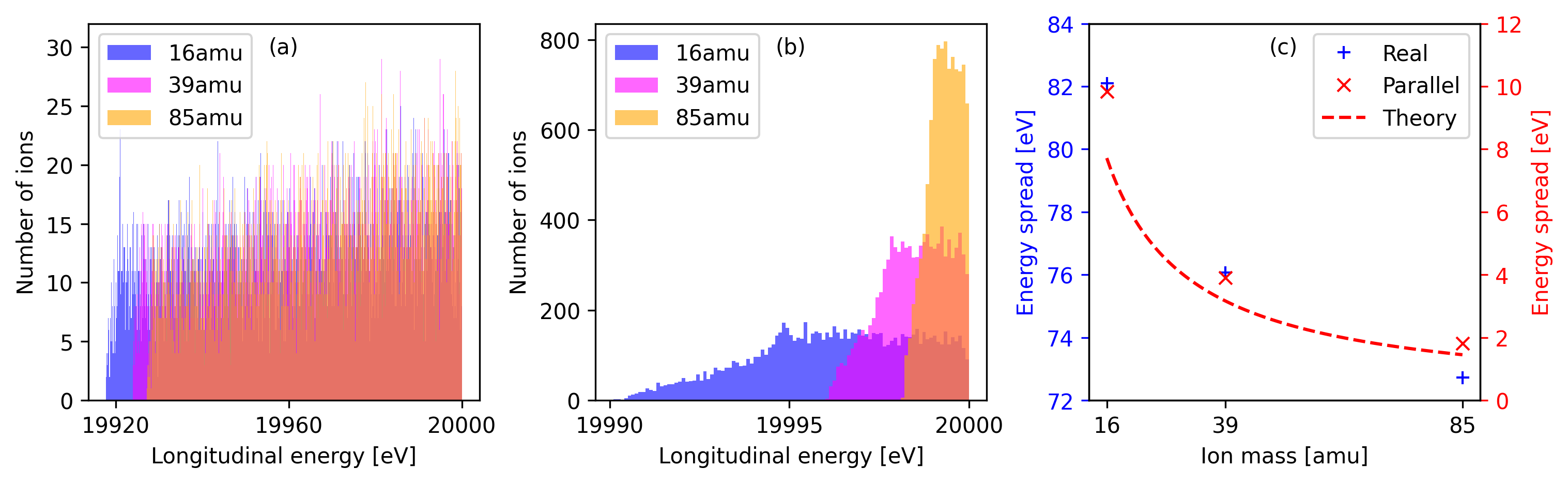}
    \caption{The longitudinal energy distribution of singly charged ion beam as a function of the ion mass for (a) the LPSC CB geometry (real), and (b) simplified extraction geometry without electrostatic focusing (parallel). The maximum energy spread of the ions in these cases is shown in (c).}
    \label{fig:meffect}
\end{figure}

 Figure~\ref{fig:meffect} shows the effect of the ion mass (for charge state 1+) on the longitudinal energy spread. The longitudinal energy spread is 73--82~\SI{}{\electronvolt} in the real extraction system but only 1.8--10~\SI{}{\electronvolt} when the electrostatic focusing effect is suppressed. The magnitude and mass dependence of the latter is well-predicted by eq.~\ref{rotation} as seen in figure~\ref{fig:meffect}(c) showing a comparison of the maximum energy spread. It is concluded that the ion mass effect is overshadowed by the electrostatic focusing effect, dominating the longitudinal energy spread.

\paragraph{Ion source potential}

\begin{figure}[htp]
    \centering
    \includegraphics[width=\textwidth]{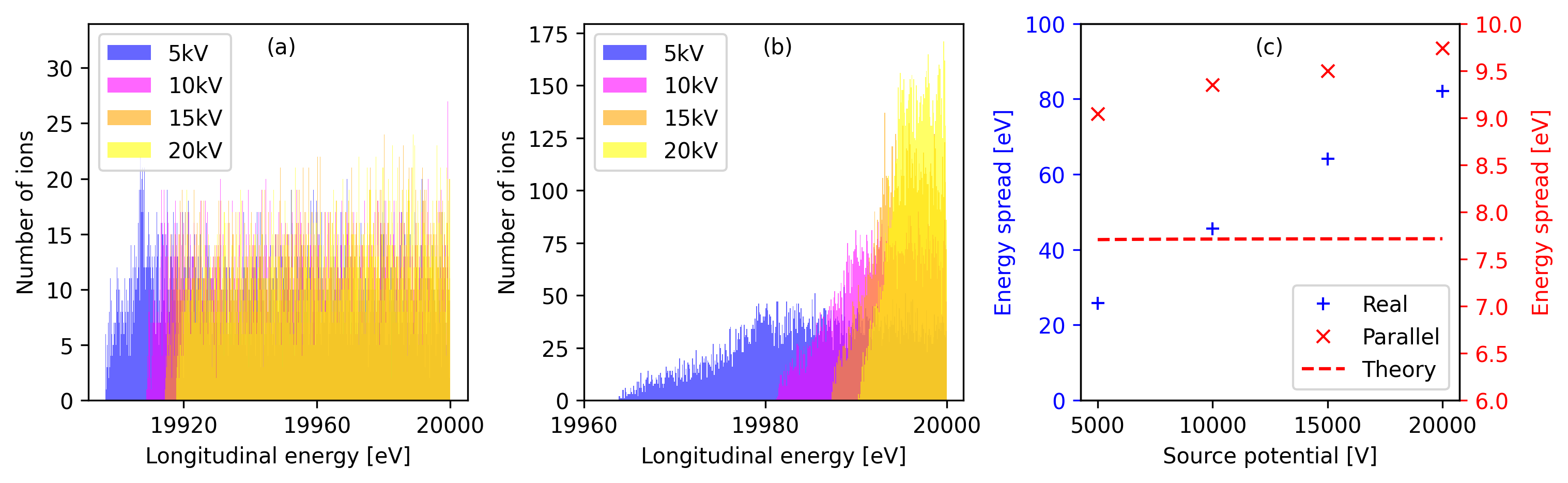}
    \caption{The longitudinal energy distribution of $^{16}$O$^{+}$ ion beam as a function of the source potential for (a) the LPSC CB geometry (real), and (b) simplified extraction geometry without electrostatic focusing (parallel). The maximum energy spread of the ions in these cases is shown in (c). The data in (a) and (b) are normalised to \SI{20}{\kilo\volt} source potential.}
    \label{fig:hv effect}
\end{figure}

Simulations of O$^+$ extraction were also carried out at ion source potentials of \SI{5}{\kilo\volt}, \SI{10}{\kilo\volt} and \SI{15}{\kilo\volt} using the real and parallel E-field models. The results are compared in figure~\ref{fig:hv effect}. In the parallel E-field extraction case, the maximum energy spread is almost constant at \SI{9}{\electronvolt} regardless of the source potential, which means that the relative energy spread $\Delta K/K$ is larger at low extraction voltages. Conversely, the energy spread scales linearly as a function of the source potential for the real extraction (note that the data in figures~\ref{fig:hv effect}(a) and \ref{fig:hv effect}(b) are normalised to \SI{20}{\kilo\volt} source potential to allow comparing the data sets), best observed in figure~\ref{fig:hv effect}(c). 

\paragraph{Ion charge state and hexapole effect}

\begin{figure}[htp]
    \centering
    \includegraphics[width=\textwidth]{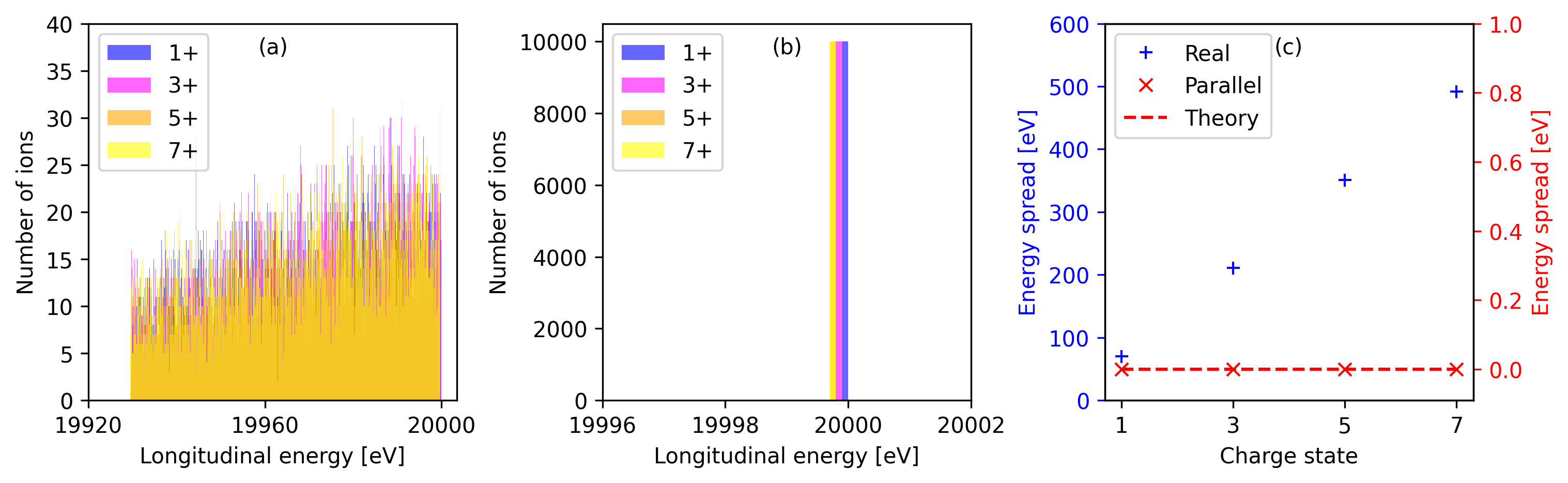}
    \caption{The longitudinal energy distribution of $^{16}$O$^{n+}$ ion beam as a function of the charge state with no magnetic field for (a) the LPSC CB geometry (real), and (b) simplified extraction geometry without electrostatic focusing (parallel). The maximum energy spread of the ions in these cases is shown in (c). The ion energy is divided by the charge state for direct comparison.}
    \label{fig:cs effect 0T}
\end{figure}

\begin{figure}[htp]
    \centering
    \includegraphics[width=\textwidth]{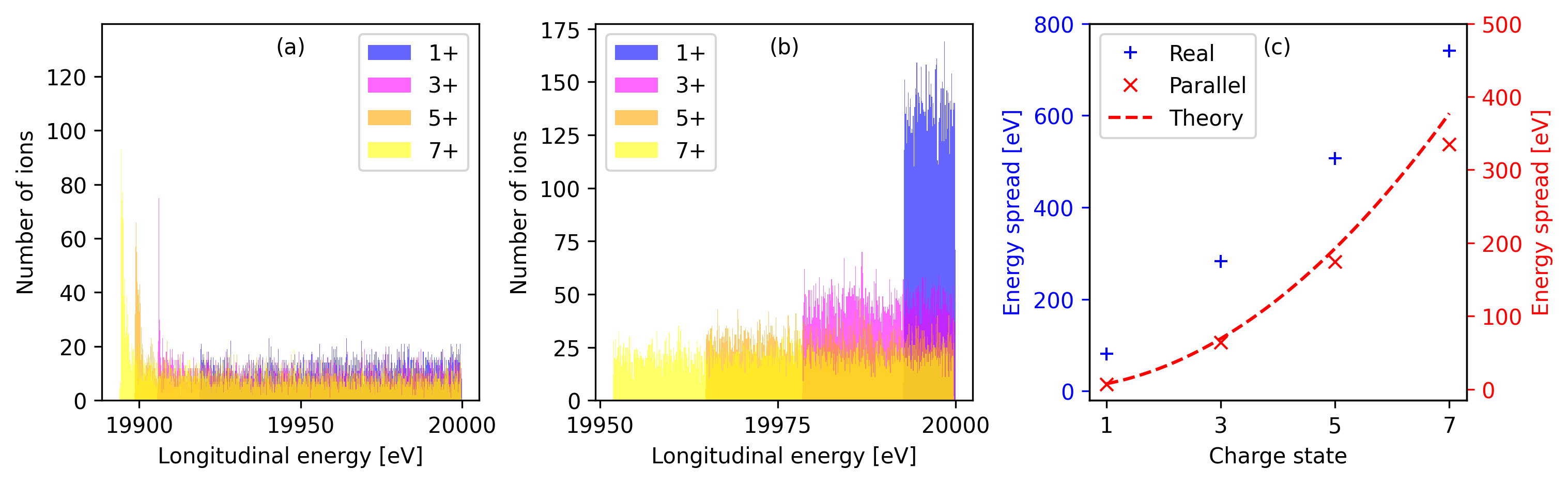}
    \caption{The longitudinal energy distribution of $^{16}$O$^{n+}$ ion beam as a function of the charge state with \SI{0.8}{\tesla} solenoid field and no hexapole field for (a) the LPSC CB geometry (real), and (b) simplified extraction geometry without electrostatic focusing (parallel). The maximum energy spread of the ions in these cases is shown in (c). The ion energy is divided by the charge state for direct comparison.}
    \label{fig:cs effect no hexa}
\end{figure}

\begin{figure}[htp]
    \centering
    \includegraphics[width=\textwidth]{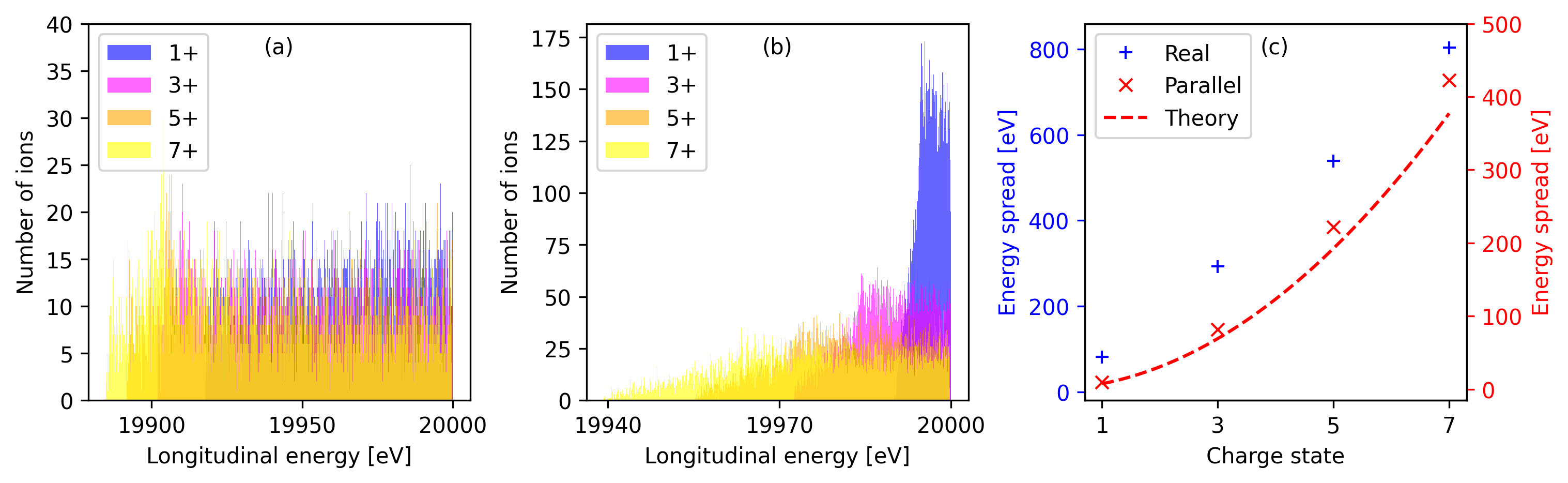}
    \caption{The longitudinal energy distribution of $^{16}$O$^{n+}$ ion beam as a function of the charge state with \SI{0.8}{\tesla} magnetic field for (a) the LPSC CB geometry (real), and (b) simplified extraction geometry without electrostatic focusing (parallel). The maximum energy spread of the ions in these cases is shown in (c). The ion energy is divided by the charge state for direct comparison.}
    \label{fig:cs effect}
\end{figure}

The effect of the charge state of oxygen on the longitudinal energy spread is depicted in figures.~\ref{fig:cs effect 0T}, \ref{fig:cs effect no hexa} and \ref{fig:cs effect} in three cases: no magnetic field, nominal \SI{0.8}{\tesla} field without the hexapole field and \SI{0.8}{\tesla} with the hexapole component (the last one being considered before as a baseline). The results obtained with charge states 3+, 5+ and 7+ have been divided by the charge state to normalize the data to \SI{20}{\kilo\electronvolt} maximum energy. The following conclusions can be made here: (i) The normalized energy spread caused by the electrostatic focusing effect is constant for all charge states, while the magnetic field induces a charge state dependent effect as predicted by eq.~\ref{rotation}. (ii) The hexapole component has a small, charge state -dependent, effect on the energy spread causing a ``tail'' towards lower longitudinal energies than with the solenoid field only, which explains the small deviation from the theoretical prediction (eq.~\ref{rotation}) in most cases studied thus far.

\paragraph{Plasma potential}

\begin{figure}[htp]
    \centering
    \includegraphics[width=\textwidth]{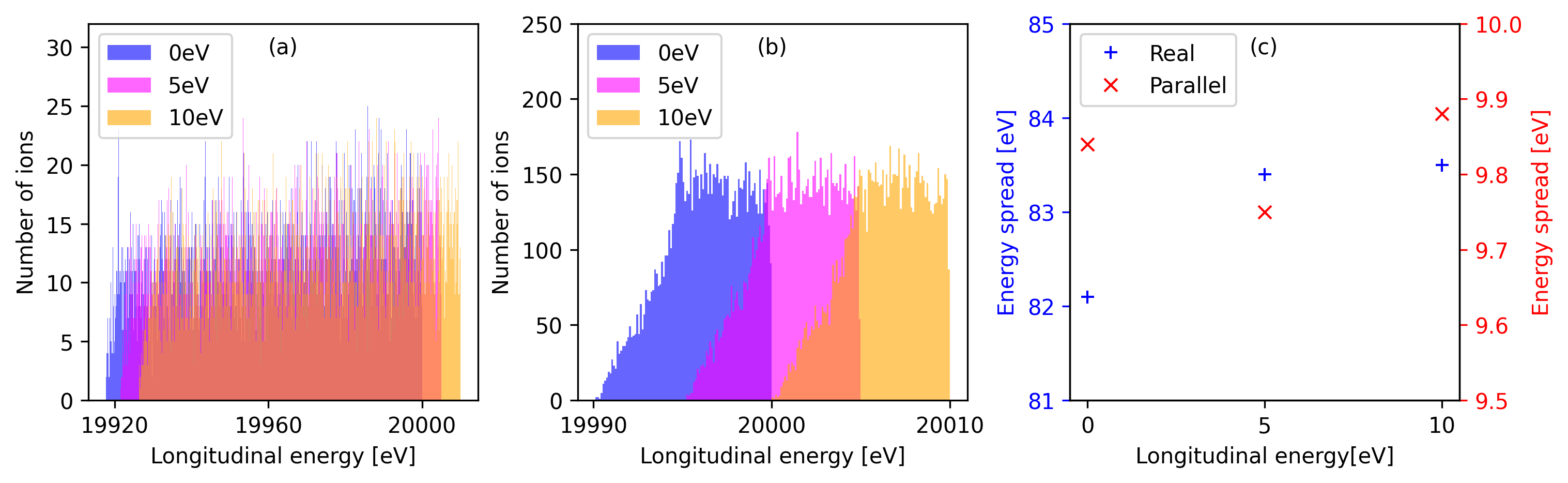}
    \caption{The longitudinal energy distribution of $^{16}$O$^{+}$ ion beam as a function of the additional longitudinal energy for (a) the LPSC CB geometry (real), and (b) simplified extraction geometry without electrostatic focusing (parallel). The maximum energy spread of the ions in these cases is shown in (c).}
    \label{fig:Ez effect}
\end{figure}

 Figure~\ref{fig:Ez effect} shows the effect of introducing an additional longitudinal energy on $^{16}$O$^+$ ions. The extra energy represents the plasma potential (up to \SI{10}{\volt}). The added longitudinal energy shifts the energy distribution by the corresponding value towards higher energies.

\begin{figure}[htp]
    \centering
    \includegraphics[width=\textwidth]{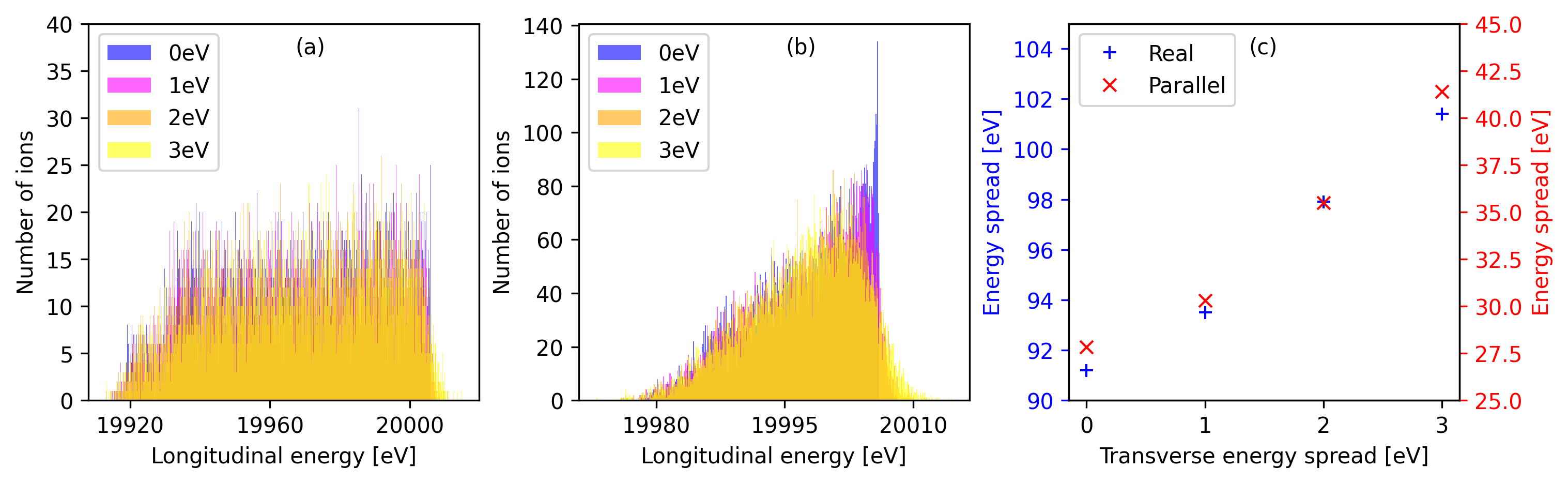}
    \caption{The longitudinal energy distribution of $^{16}$O$^{+}$ ion beam as a function of the initial transverse energy spread for (a) the LPSC CB geometry (real), and (b) simplified extraction geometry without electrostatic focusing (parallel). The maximum energy spread of the ions in these cases is shown in (c).}
    \label{fig:Eyspread effect}
\end{figure}

\paragraph{Initial transverse energy spread (ion temperature)}
The effect of the initial transverse energy spread on the longitudinal ion energy at the end of the simulation domain is illustrated in figure~\ref{fig:Eyspread effect}. 
The ions were launched in a random transverse direction with a \SI{6}{\electronvolt} mean energy and energy spread values of \SI{0}{\electronvolt}, \SI{1}{\electronvolt}, \SI{2}{\electronvolt} and \SI{3}{\electronvolt} depicting the ion temperature. The excess longitudinal energy up to \SI{10}{\electronvolt}, which is smaller than the (up to) \SI{90}{\electronvolt} energy spread caused by the electrostatic focusing and magnetic field effects at longitudinal energies below \SI{20}{\kilo\electronvolt}, can be clearly observed. The excess energy implies that initial transverse energy is converted into longitudinal energy as predicted by eq.~\ref{moment}. Besides the plasma potential, the conversion of transverse energy into longitudinal energy is the only process emerging at longitudinal energies higher than \SI{20}{\kilo\electronvolt}.

\begin{figure}[htp]
    \centering
    \includegraphics[width=\textwidth]{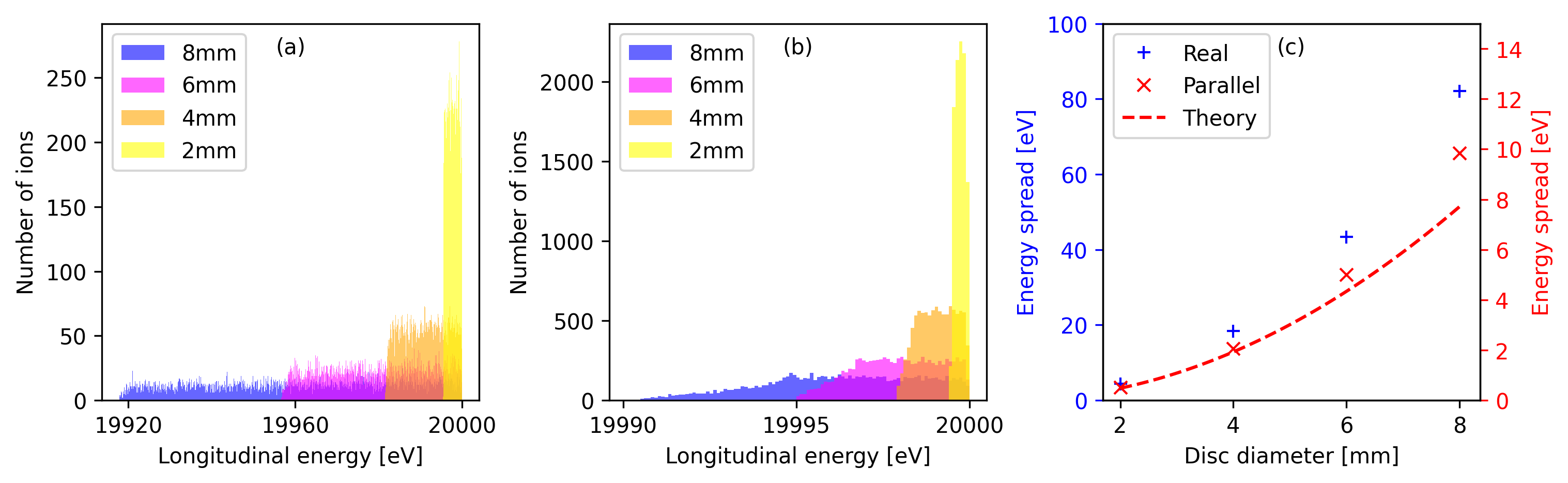}
    \caption{The longitudinal energy distribution of $^{16}$O$^{+}$ ion beam as a function of the plasma electrode disc aperture for (a) the LPSC CB geometry (real), and (b) simplified extraction geometry without electrostatic focusing (parallel). The maximum energy spread of the ions in these cases is shown in (c).}
    \label{fig:disc effect}
\end{figure}

\paragraph{Effective extraction radius (spatial distribution of ions)}
Figure~\ref{fig:disc effect} shows the effect of the plasma electrode aperture diameter (launch disk diameter) on the longitudinal energy spread of O$^+$ ions at the end of the simulation domain. The results clearly demonstrate that the longitudinal energy spread depends strongly on the effective extraction radius of the ions in both cases. This is because the electrostatic focusing effect induces lower transverse energies to those ions near the optical axis and, on the other hand, the effect of the magnetic field induced rotation scales with the radius squared (see eq.~\ref{rotation}). 

\begin{figure}[htp]
    \centering
    \includegraphics[width=\textwidth]{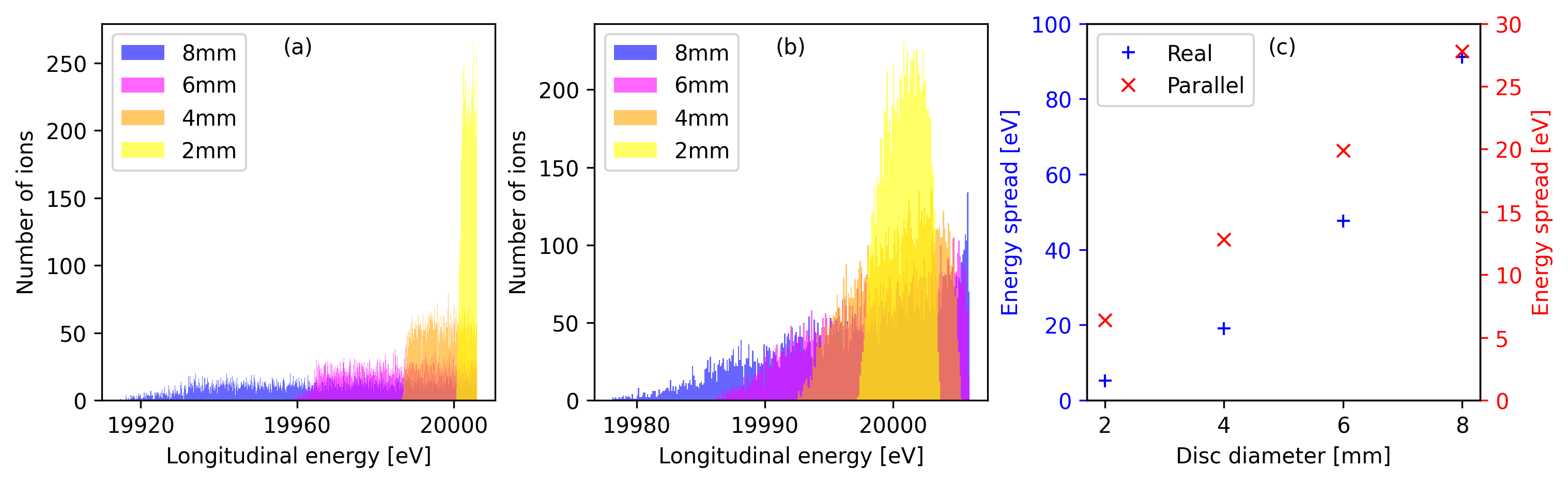}
    \caption{The longitudinal energy distribution of $^{16}$O$^{+}$ ion beam as a function of the plasma electrode disc aperture with (isotropic) initial transverse energy of \SI{6}{\electronvolt} and no initial transverse energy spread. (a) is for the LPSC CB geometry (real), and (b) for the simplified extraction geometry without electrostatic focusing (parallel). The maximum energy spread of the ions in these cases is shown in (c).}
    \label{fig:disc effect ET6 ETS0}
\end{figure}

\begin{figure}[htp]
    \centering
    \includegraphics[width=\textwidth]{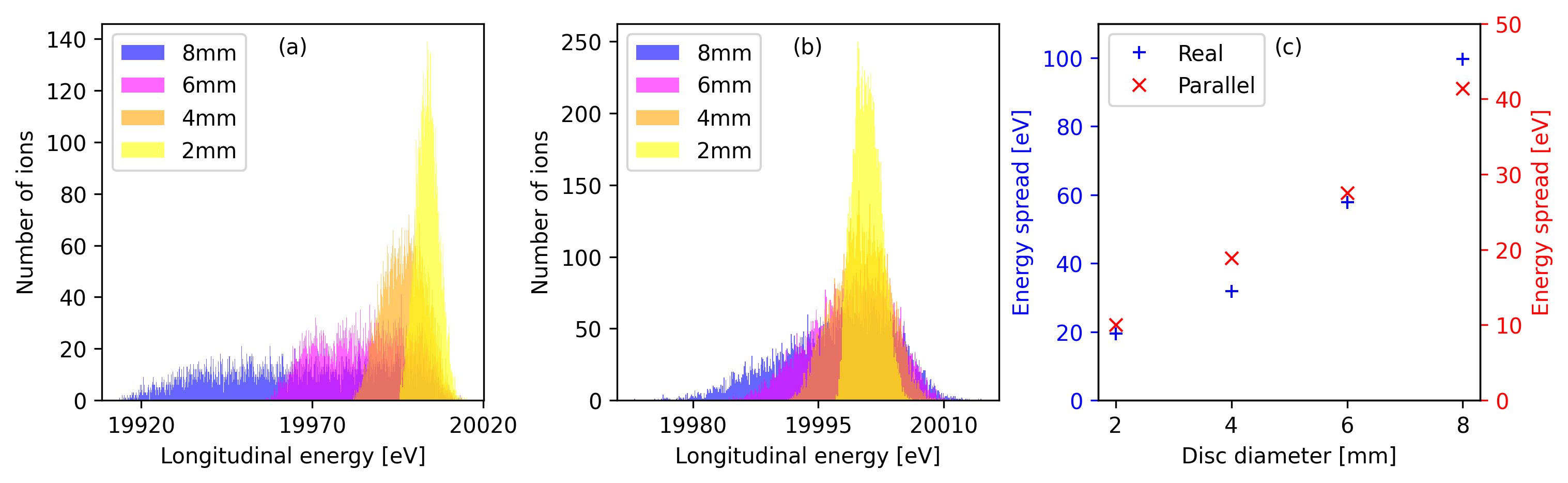}
    \caption{The longitudinal energy distribution of $^{16}$O$^{+}$ ion beam as a function of the plasma electrode disc aperture with initial transverse energy of \SI{6}{\electronvolt} and \SI{3}{\electronvolt} initial transverse energy spread. (a) is for the LPSC CB geometry (real), and (b) for the simplified extraction geometry without electrostatic focusing (parallel). The maximum energy spread of the ions in these cases is shown in (c).}
    \label{fig:disc effect ET6 ETS3}
\end{figure}

The simulations were repeated with (i) an initial transverse energy of \SI{6}{\electronvolt} with no energy spread, see figure~\ref{fig:disc effect ET6 ETS0}, together with (ii) an initial transverse energy of \SI{6}{\electronvolt} and a \SI{3}{\electronvolt} transverse energy spread, see figure~\ref{fig:disc effect ET6 ETS3}. The other parameters were kept unchanged. This was to confirm that the conversion of the transverse energy to longitudinal energy does not depend on the radial coordinate of the ions at the plasma-beam boundary replicated by the launching disk. The conclusions is that the electrostatic focusing and rotation of the beam induced by the decreasing magnetic field cause the longitudinal ion energies to spread towards lower values. Both these contributions depend on the radial coordinate of the ions contrary to the process converting initial transverse energy to longitudinal energy, which occurs independent of the radial coordinate. 

The presented simulation results imply that the diagnostics of the longitudinal energy spread of the ion beams extracted from an ECRIS requires capturing the whole beam for the measurement. That is because the energy spread of the ions depends strongly on their initial radial coordinate at the plasma-beam boundary. On the other hand, studying the longitudinal energy spread related to the ECRIS plasma conditions (not electrostatic focusing or magnetic field effects) requires selecting the ions closest to the beam axis as discussed further in the following sections.

\subsection{\textsc{IBSimu} extraction simulations}

To complement the \textsc{Simion} simulations we used \textsc{IBSimu} to study the effects of plasma sheath and space charge on the longitudinal energy spread of the extracted ion beam. This was done for three cases: \textbf{(i)} without plasma or space charge to compare and cross-validate the simulation results of \textsc{IBSimu} and \textsc{Simion}, \textbf{(ii)} with space charge but no plasma extraction model, i.e.\ emitting disc similar to \textsc{Simion}, and \textbf{(iii)} with plasma (meniscus) and space charge. Comparison of the results of the latter two allows independently estimating the effects of plasma meniscus formation and space charge on the longitudinal energy spread of the ion beam.

The simulations presented in this section used the same reference case parameters, which were used with \textsc{Simion} (see table~\ref{tab:reference}). The first and second model have a geometry equivalent to the first \textsc{Simion} model with the artificial HV plate at the plasma electrode aperture. The plasma parameters used as inputs for the third model are: \SI{10}{\electronvolt} plasma temperature $T_e$, \SI{20}{\volt} plasma potential, and initial ion energy of $E_0 = T_e/2 =$ \SI{5}{\electronvolt} according to Bohm theory~\cite{Bohm}. The ion temperature in the plasma was set to zero. 

Most of the simulations done with \textsc{Simion} were first replicated with \textsc{IBSimu} without the plasma model or space charge. Generally the longitudinal energy spreads obtained from the simulations agree well with each other -- \textsc{IBSimu} and \textsc{Simion} produced practically the same results. For example, in the reference case the difference of the energy spread is 6 \%. The minor differences are presumably explained by small differences in the simulations, for example the mesh size used, number of particles simulated and electric field artefacts near the circular plasma aperture boundary.

\begin{figure}[hbt]
    \centering
    \includegraphics[width=\textwidth]{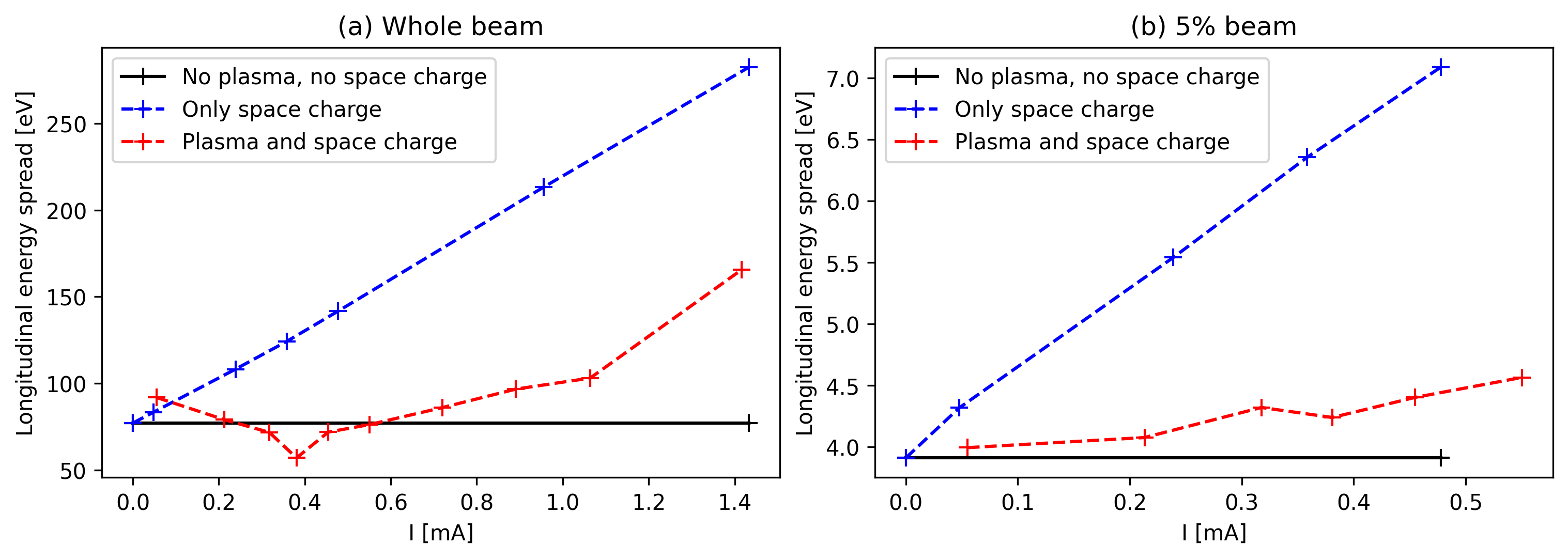}
    \caption{The longitudinal energy spread of $^{16}$O$^+$ ion beam as a function of beam current using \textsc{IBSimu}. All the energy spreads are calculated at the end of simulation region at $z = $ \SI{1000}{\milli \meter}. (a) With the whole beam and (b) with 5\% of the ions closest to the beam axis.}
    \label{fig:IBSimu_dE_vs_I}
\end{figure}

Figure~\ref{fig:IBSimu_dE_vs_I}(a) shows the longitudinal energy spread at the end of the simulation domain (\SI{1000}{\milli \meter}) as a function of the ion beam current, with the three models. The energy spread in the first model (without plasma or space charge) is constant at about \SI{77}{\electronvolt} independent of the beam current as expected due to the lack of the space charge forces -- in other words, the beam current is effectively zero in this model. Only the source magnetic field and the electric field defined by the extraction electrodes affect energy spread of the beam. With the second model, taking in account the space charge but not plasma effects, the increase of longitudinal energy spread with the beam current is almost linear, which is caused by the growing space charge forces of the beam increasing the beam divergence. On the second model, electrostatic focusing and extraction magnetic field are the dominant effects at beam currents less than \SI{0.5}{\milli \ampere} (typical for CB ion sources) causing the majority of the longitudinal energy spread. Space charge effects become dominant at currents higher than this. In the third model, which is the most realistic, the beam divergence and, thus, the longitudinal energy spread are strongly affected by the focusing properties of the plasma sheath. The plasma focusing is also coupled with the electrostatic focusing provided by the electrodes and the space charge effects on the following beam transport. The plasma effects cause the longitudinal energy spread to be lower than in the case where only space charge effects are taken in account. 

Figure~\ref{fig:IBSimu_dE_vs_I}(b) illustrates the longitudinal energy spread dependence on the total beam current for 5\% of the ions closest to the beam axis. The same qualitative behaviour as in figure~\ref{fig:IBSimu_dE_vs_I}(a) can be seen but the absolute energy spread is greatly reduced due to the transverse component of the electric field being very small near the beam axis. This has great relevance in probing the plasma potential induced energy spread with a retarding field analyser as elaborated further in the following sections. In essence, studying the plasma conditions requires significant collimation of the particle beam as the energy spread of the whole beam depends heavily on the space charge and plasma meniscus effects. In the third model, which is the most realistic, we see (almost) a linear rise in longitudinal energy spread with the beam current but the absolute change of the longitudinal energy spread is only \SI{0.5}{\electronvolt}. This highlights the fact that if the beam is collimated at the measurement device, the space charge and plasma meniscus effects on the longitudinal energy spread are suppressed.

\begin{figure}[hbt]
    \centering
    \includegraphics[width = \textwidth]{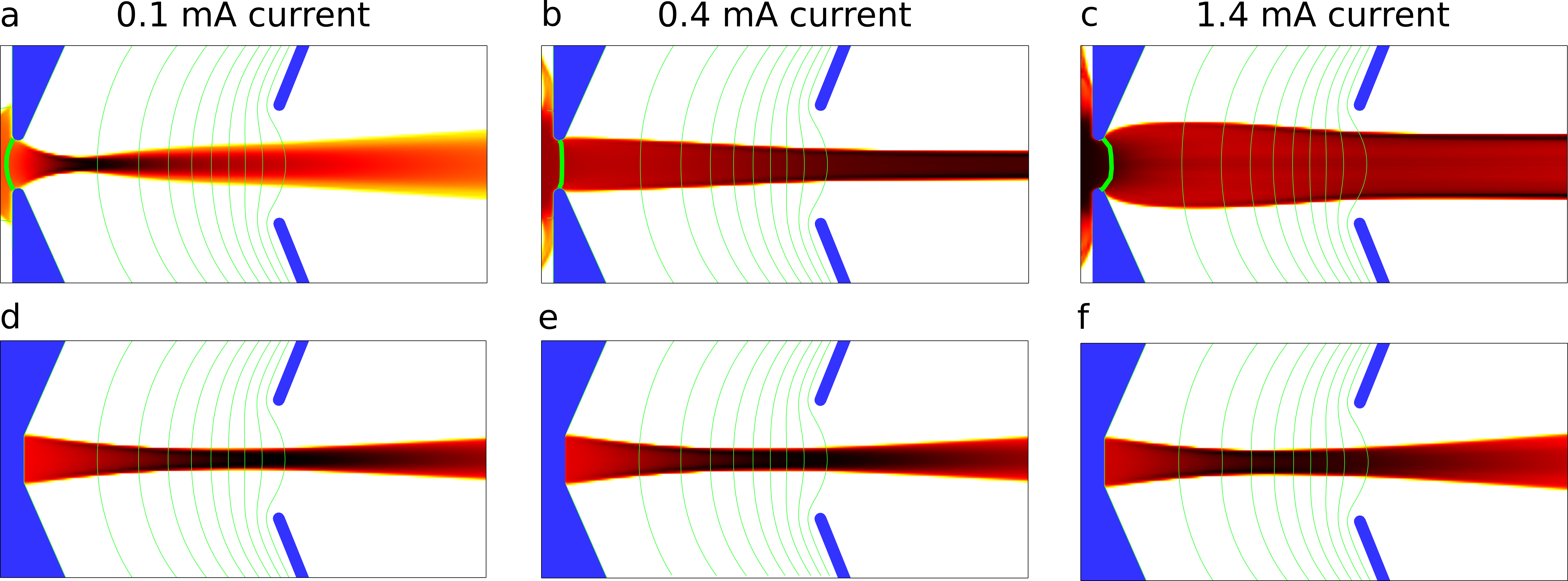}
    \caption{\textsc{IBSimu} simulations with \SI{0.1}{\milli \ampere}, \SI{0.4}{\milli \ampere} and \SI{1.4}{\milli \ampere} beam currents. The top row (a,b and c) simulations are with plasma and space charge, and the plasma meniscus is highlighted with a green arc. The plasma meniscus is defined as the potential surface where the electric potential equals the plasma chamber potential. Increasing the beam current changes the shape of plasma meniscus from concave to straight (slightly convex) to convex. The bottom row (d,e and f) simulations are with space charge but with no plasma so the beam starts from a constant potential plane at the location of the extraction aperture.}
    \label{fig:extraction_cartoon}
\end{figure}

Figure~\ref{fig:extraction_cartoon} shows examples of the simulations, which are used to interpret the results shown in figure~\ref{fig:IBSimu_dE_vs_I} (a). \textit{The low current} case (\SI{0.1}{\milli \ampere}) is the only one where the longitudinal energy spread is higher with the plasma model than with the other two models. Figure~\ref{fig:extraction_cartoon}(a) shows the overfocusing of the beam due to the concave plasma sheath, which causes high spatial concentration of charges at the crossover point and hence strong space charge forces. These lead to high transverse and longitudinal energy spreads compared to the other models.  Without the plasma (figure~\ref{fig:extraction_cartoon}(d)) the beam focus caused by the electrode geometry is much weaker resulting in a smaller energy spread.

With \textit{medium current} (\SI{0.4}{\milli \ampere}) the plasma sheath is nearly flat and the beam is parallel at the plasma-beam boundary. The focusing force of the electrode geometry is partially counteracted by the space charge force resulting in the beam remaining parallel or slightly converging as shown in figure~\ref{fig:extraction_cartoon}(b). The divergence of the beam increases slightly after the extraction due to the space charge forces leading to a small longitudinal energy spread. Without the plasma the focusing effect caused by the electrode geometry is stronger than with the plasma, which leads to a larger energy spread.

In the \textit{high current case} (\SI{1.4}{\milli \ampere}) the beam diverges at the start due to the convex plasma sheath, which is counteracted by the focusing effect of the extraction field resulting in a parallel beam as shown in figure~\ref{fig:extraction_cartoon}(c). However, due to the high charge density (high current) of the beam, the space charge effects start to increase the transverse energy of the beam causing a spread in the longitudinal energy as well. With the no-plasma-case (figure~\ref{fig:extraction_cartoon}(f)) the beam starts with much narrower spatial distribution, which again leads to high space charge forces and large energy spread. 

The \textsc{IBSimu} results can be summarized in three points: (1) at high currents the high initial divergence caused by the convex plasma sheath lowers the energy spread by mitigating the effect of the space charge forces, (2) at low currents the overfocus by the plasma sheath introduces a crossover point and high space charge forces leading into large energy spreads, (3) to study the plasma conditions the beam needs to be collimated as the magnitude of space charge and plasma meniscus effects are reduced in the collimated beam. It is worth noting that the electrostatic radial focusing arising from the extraction electrode geometry can balance the ion beam space charge effect and, thus, decrease the longitudinal energy spread, first described by Pierce~\cite{Pierce}. The plasma sheath focusing and beam space charge forces could contribute to the phenomena reported in ref.~\cite{Higurashi} where the increase of the extraction magnetic field led to significant decrease of the transverse emittance of high charge state uranium beams, contrary to the expectation. In the light of the present study, it could be that increasing the magnetic field, which typically leads to lower beam currents implying a less convex meniscus, lowers the beam divergence and transverse emittance.

\section{Retarding field analyser simulations} 
\label{RFA_simulations}

The beam extraction simulations presented above highlight the importance of appropriate selection of ions for measurements probing either the energy spread of the complete ion beam (electrostatic and magnetic field effects) without collimation vs. the energy spread caused by plasma conditions (plasma potential, ion temperature) with well-collimated beams. The energy spread measurements are typically carried out with Retarding Field Analysers (RFA)~\cite{Bibinov}, which motivates reproducing their predicted IV-curves at the end of the simulation domain for various fractions of the captured beam as described hereafter. 

\subsection{Operating principle of a retarding field analyser}

Planar retarding field analyser is an instrument commonly used for the measurement of the ion source plasma potential and the energy spread of a given ion beam species. Such device was used in the CB ECRIS context at GANIL to study the influence of the injected beam energy spread on the 1+ ion capture and charge breeding efficiency~\cite{Maunoury}. At JYFL (University of Jyvaskyla) Accelerator Laboratory, an RFA was developed to estimate the maximum plasma potential value of ECR ion sources~\cite{JYFL_RFA}. One version of this RFA was installed at the LPSC N+ beam line in 2009, see figure~\ref{fig:1+N+ test bench}, to estimate the impact of the injected 1+ ions on the plasma potential~\cite{Lamy2010}. The JYFL RFA was further developed as a part of this study.

\begin{figure}[H]
    \centering
    \includegraphics[width=\textwidth]{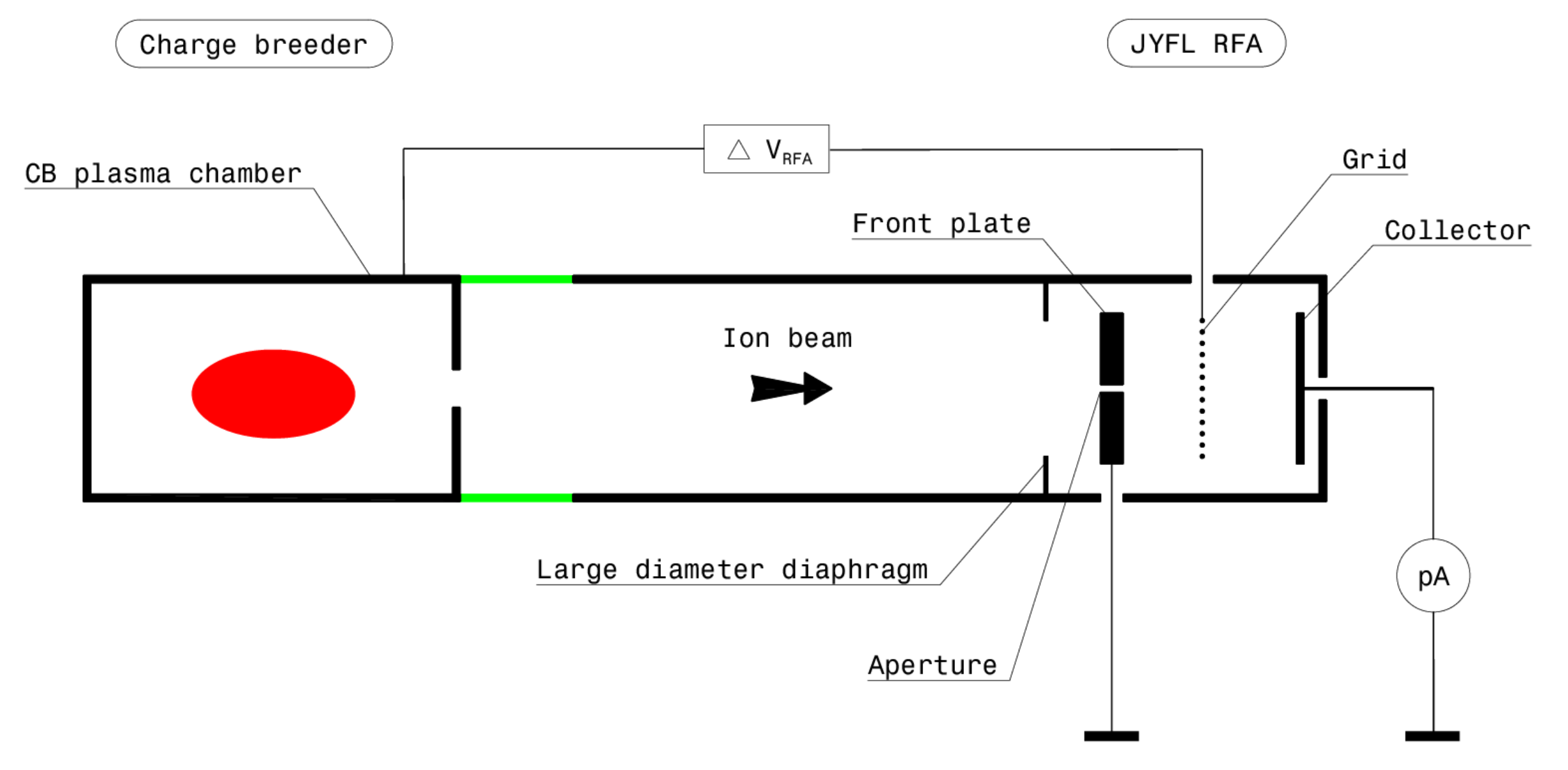}
    \caption{Schematic of the retarding field analyser setup installed on the 1+N+ test bench at LPSC.}
    \label{fig:JYFL RFA}
\end{figure}

In general, RFAs rely on the generation of a variable electrostatic potential on the beam axis to filter the ions as a function of their energy. It is composed of (i) a collimating aperture to select the ions close to the beam axis, (ii) a retarding grid to create the electrostatic potential and (iii) a collector electrode, connected to ground via an ammeter, measuring the current of the ions that pass through the grid. The ions are decelerated between the collimator and the grid, and re-accelerated back to their original energy between the grid and the collector. To prevent secondary electron emission from the collector surface, a grid with negative potential can be placed upstream from the collector. Depending on the geometry, the electrostatic potential configuration can be improved by a grounded grid placed just after the collimator. Alternatively, the grid can be replaced by a cylindrical focusing electrode, set between the collimating aperture and the retarding grid. Figure~\ref{fig:JYFL RFA} shows a schematic of the experimental setup used for the measurement of the energy spread of the ion beam extracted from an ECRIS. The RFA grid is connected to a power supply floating at the CB ECRIS potential $V_s$. The RFA (grid) voltage, $\Delta V_{\textrm{RFA}}$ is swept from negative to positive voltage (some tens of volts) to record an IV-curve representing the fraction of ions with longitudinal energies higher than $q(V_s+\Delta V_{\textrm{RFA}})$. The energy distribution function is obtained by taking the absolute value of the derivative of the IV-curve.

\begin{figure}[H]
    \centering
    \includegraphics[width=\textwidth]{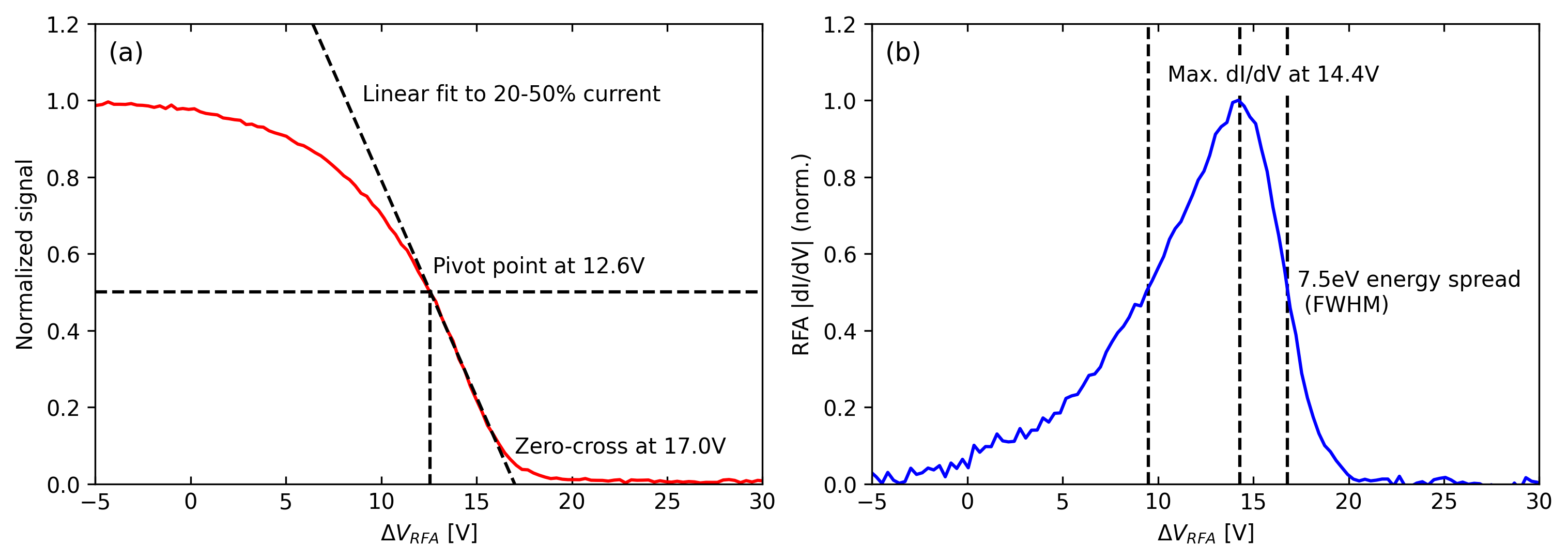}
    \caption{An example RFA IV-curve (a) and its derivative (b), measured with an O$^+$ ion beam extracted from the ECRIS CB at LPSC. Various methods -- the pivot point, zero-crossing of the linear fit and maximum of the derivative -- to determine the plasma potential are illustrated.}
    \label{fig:RFA example}
\end{figure}

Figure~\ref{fig:RFA example}(a) shows an example of an IV-curve measured with the (modified) RFA at LPSC. The corresponding derivative is shown in figure~\ref{fig:RFA example}(b). The figures illustrate different methods that have been used in the literature to estimate the plasma potential and energy spread from the RFA-data. Most often the maximum of the IV-curve derivative, i.e.\ \SI{14.4}{\volt} in the example, is interpreted as the plasma potential (see e.g. ref.~\cite{Gahan} for comparison to Langmuir-probe IV-curve interpretation). However, this method ignores the fact that, especially in ECRIS plasmas, the ions are collisional and the length scale of the electrostatic potential variation is longer than the mean free path of the ions, which results in a distribution of energies not necessarily peaked at the plasma potential. Detailed explanation of the plasma sheath effects in RFA data interpretation can be found from the literature~\cite{Fisher}. In ref.~\cite{JYFL_RFA} this ambiguity was avoided by defining the plasma potential to correspond to the maximum energy of the ions, determined from the zero-crossing of a linear fit to the normalised IV-curve made between the 20--50\% values. In the given example the plasma potential corresponding to such method is \SI{17.0}{\volt}. Here we will demonstrate that, according to simulations taking into account the afore-mentioned processes inducing the energy spread of the ion beams extracted from an ECRIS, the best indicator for the plasma potential is the 50\% "pivot point" of the IV-curve measured with significant collimation, i.e.\  \SI{12.6}{\volt} in figure~\ref{fig:RFA example}, while further characteristics of the IV-curve can be attributed to the ``effective ion temperature'' of the beam.    

\subsection{RFA IV-curve simulations}\label{RFA_simulations}

A comprehensive set of \textsc{Simion} simulations were carried out to quantify the effects of the ion source extraction parameters on the beam energy spread and the corresponding synthetic IV-curves of the RFA. In particular, we demonstrate the effect of collimating the beam by selecting 2\%, 14\% and 100\% of the simulated beam particles closest to the beam axis for energy spread analysis at various ion source parameter combinations. The first two beam fractions account for the effects of RFA entrance collimators used in our experiments (described later), whereas the 100\% fraction represents the IV-curve and energy spread of the total ion beam, relevant for matching the beam into the acceptance of an accelerator. The \textsc{Simion} simulations were carried out with the real extraction model using the \SI{8}{\milli\meter} diameter emitter disc representing the extraction aperture. The reference parameters are summarized in table \ref{tab:reference RFA}
and the influence of each of these parameters on the RFA-IV curves was studied individually.

\begin{table}[H]
    \caption{Reference parameters of the RFA simulations.}
    \centering
    \begin{tabular}{|c|c|}
    \hline
    Ion mass (amu)            & 16 \\
    Ion charge (e)          & 1 \\
    B-field at extraction (\SI{}{\tesla})  & 0.8 \\
    Ion source potential (\SI{}{\kilo\volt})  & 20 \\
    Initial ion energy (\SI{}{\electronvolt})     & 6 \\
    Ion temperature (\SI{}{\electronvolt})    & 2 \\
    \hline
    \end{tabular}
    \label{tab:reference RFA}
\end{table}

The results are presented in the form of the predicted RFA IV-curves at the end of the simulation domain, i.e.\ taking the integral of the energy distribution while considering the stopping of the ions by the RFA potential. As such, the synthetic IV-curves are plotted as a function of the potential difference between the RFA and the ion source, $\Delta V_{RFA}$. It is worth emphasising that the simulated RFA IV-curves are ideal in the sense that they do not take into account instrumental effects, e.g.\ energy resolution of the RFA. The integral of the initial energy distributions are shown as dashed lines for the comparison with the final distributions.

\begin{figure}[htp]
    \centering
    \includegraphics[width=\textwidth]{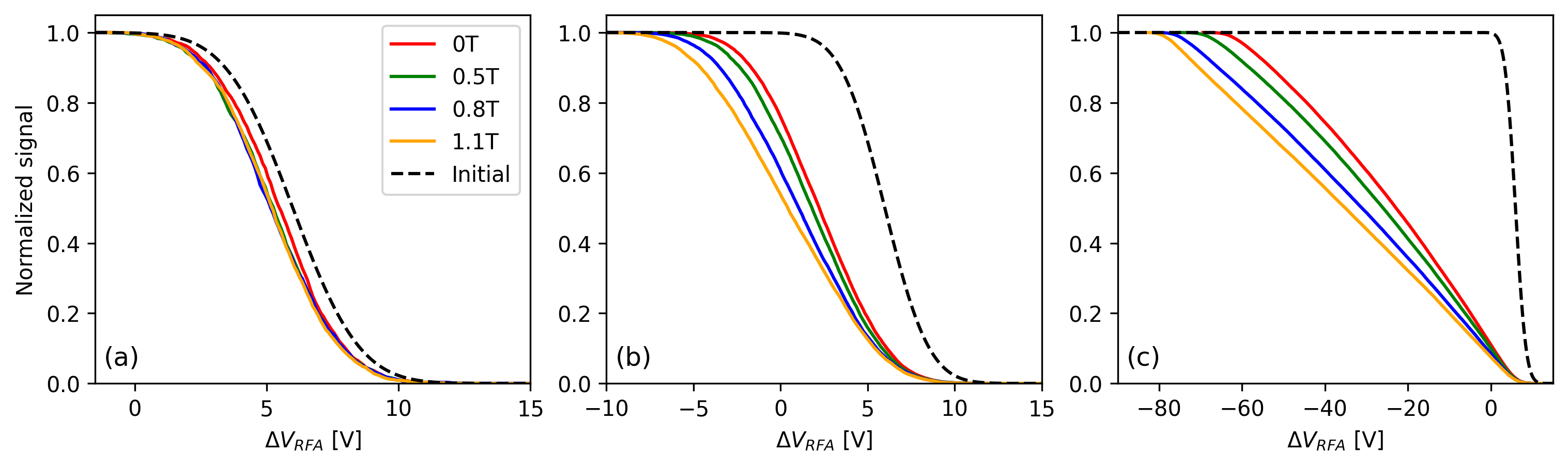}
    \caption{ Simulated RFA IV-curves of O$^+$ ions for different CB solenoidal magnetic fields with (a) 2\% (b) 14\% and (c) 100\% fraction of the ions the closest to the beam axis considered. Note different horizontal scales. }
    \label{fig:B effect}
\end{figure}

\paragraph{Magnetic field}
First, we varied the axial magnetic field strength of the ion source to study its effect on the longitudinal energy distribution of the ion beam, keeping the other parameters listed in table \ref{tab:reference RFA} unchanged. Figure~\ref{fig:B effect} presents the simulated RFA IV-curves for the 2\% (a), 14\% (b) and 100\% (c) beam fractions closest to the axis with the magnetic field maximum ranging from \SI{0}{\tesla} to \SI{1.1}{\tesla}. In the 2\% case, the final distributions are in good agreement with the initial one whereas for larger fractions of the beam, the electrostatic focusing effect dominates and the RFA IV-curves deviate significantly from the initial distribution.

\begin{table}[H]
    \begin{center}
    \begin{tabular}{|c|c c c c|} 
    \hline
    Magnetic field (\SI{}{\tesla}) & 0 & 0.5 & 0.8 & 1.1 \\ [0.5ex] 
     \hline
    Pivot point 2\% fraction (\SI{}{\volt}) & 5.5 & 5.3 & 5.2 & 5.2 \\ 
    Pivot point 14\% fraction  (\SI{}{\volt}) & 2.2 & 1.8 & 1.0 & 0.4 \\
    Pivot point 100\% fraction  (\SI{}{\volt}) & $-$22.9  & $-$26.0 & $-$31.1 & $-$35.1  \\ 
    \hline
    Max dI/dV 2\% fraction (\SI{}{\volt}) & 5.1 & 5.5 & 4.6 & 5.3 \\ 
    Max dI/dV 14\% fraction  (\SI{}{\volt}) & 2.3 & 1.2 & 1.3 & 0.3 \\
    Max dI/dV 100\% fraction  (\SI{}{\volt}) & $-$3.8  & $-$1.2 & 0.1 & $-$12.6  \\ 
    \hline
    Zero cross 2\% fraction (\SI{}{\volt}) & 8.1 & 8.1 & 8.1 & 8.0 \\ 
    Zero cross 14\% fraction  (\SI{}{\volt}) & 6.6 & 6.3 & 6.2 & 6.3 \\
    Zero cross 100\% fraction  (\SI{}{\volt}) & 7.1 & 7.0 & 7.0 & 6.8  \\ 
    \hline    
    \end{tabular}
    \end{center}
    \caption{The values of the pivot point, maximum of dI/dV and IV-curve linear fit zero-crossing as a function of the solenoid field strength.}    
    \label{tab:B effect}
\end{table}

Table~\ref{tab:B effect} summarises the three methods -- pivot point, maximum of the IV-curve derivative and zero-crossing of the linear fit -- used to determine the plasma potential from the RFA data with varying fraction of the beam. The pivot point is found at about \SI{0.8}{\volt} below the average initial longitudinal energy defined by the plasma potential with extraction field $>$\SI{0.5}{T}, which is typical for the CB ECRIS. This shift is due to the transverse energy component that still exists even for the ion population closest to the beam axis. In the 14\% and 100\% cases, the initial and final energy distributions differ significantly with the energy spread increasing as more ions are considered. These results imply that the effects of the electrostatic focusing and magnetic field induced rotation can be suppressed in the experiment with appropriate collimation, selecting only a few percent of the ions nearest to the beam axis for longitudinal energy analysis.  It is worth noting that considering the pivot point or the maximum of the IV-curve derivative deviate significantly (even yielding unphysical negative values) from the plasma potential if a large fraction of the ions is detected at the RFA collector. In that case the zero-crossing of the linear fit imposed on the IV-curve yields the most accurate estimate as argued in ref.~\cite{JYFL_RFA}.

\begin{figure}[H]
    \centering
    \includegraphics[width=\textwidth]{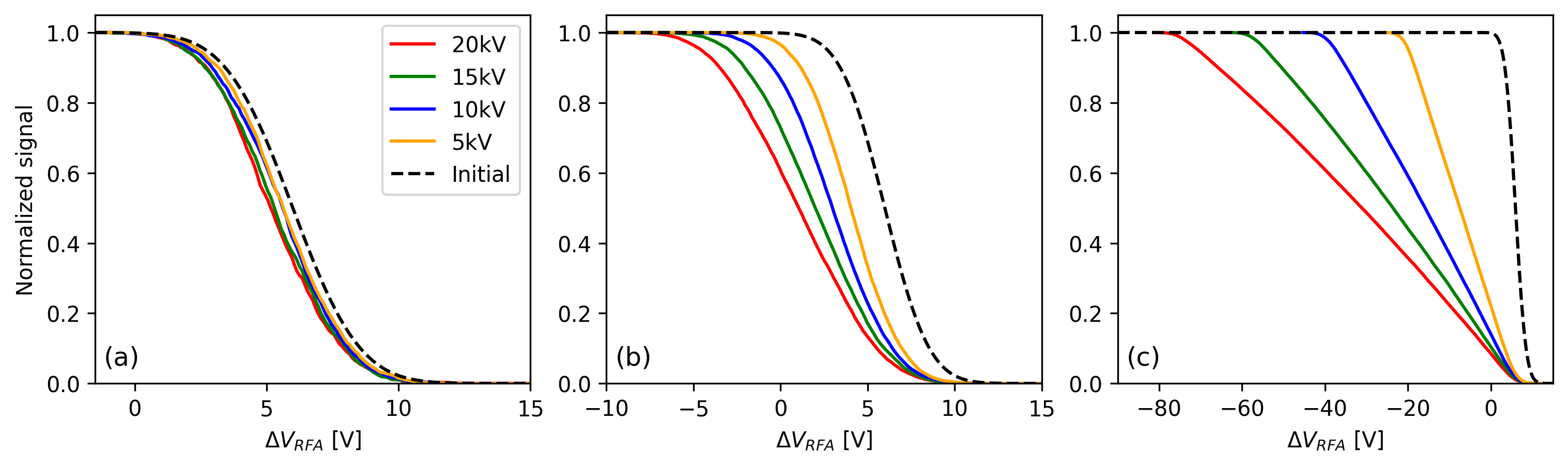}
    \caption{ Simulated RFA IV-curves of O$^+$ ions for different source potentials with (a) 2\% (b) 14\% and (c) 100\% fraction of ions closest to the beam axis considered.}
    \label{fig:Vs effect}
\end{figure}

\paragraph{Source potential}
Simulations were done for source potentials of 5, 10, 15 and \SI{20}{\kilo\volt}, the other parameters being set at their reference values. Figure~\ref{fig:Vs effect} presents the predicted IV-curves for the 2\% (a), 14\% (b) and 100\% (c) fractions of ions closest to the beam axis.

A good agreement between the initial and final distributions is obtained only for the 2\% beam fraction. When the fraction of the collected ions increases, the RFA IV-curves start deviating from the initial distribution. The effect is pronounced at high source potentials, where the conversion of longitudinal energy to transverse energy due to the electrostatic focusing causes a larger spread of the longitudinal energy (see figure~\ref{fig:hv effect}(c) for comparison). This implies that RFA measurements should be performed at lowest possible source potential if the collimation of the beam is insufficient. On the other hand, if the beam is heavily collimated, the source potential does not affect the interpretation of the RFA data.

\begin{figure}[H]
     \centering
      \includegraphics[width=\textwidth]{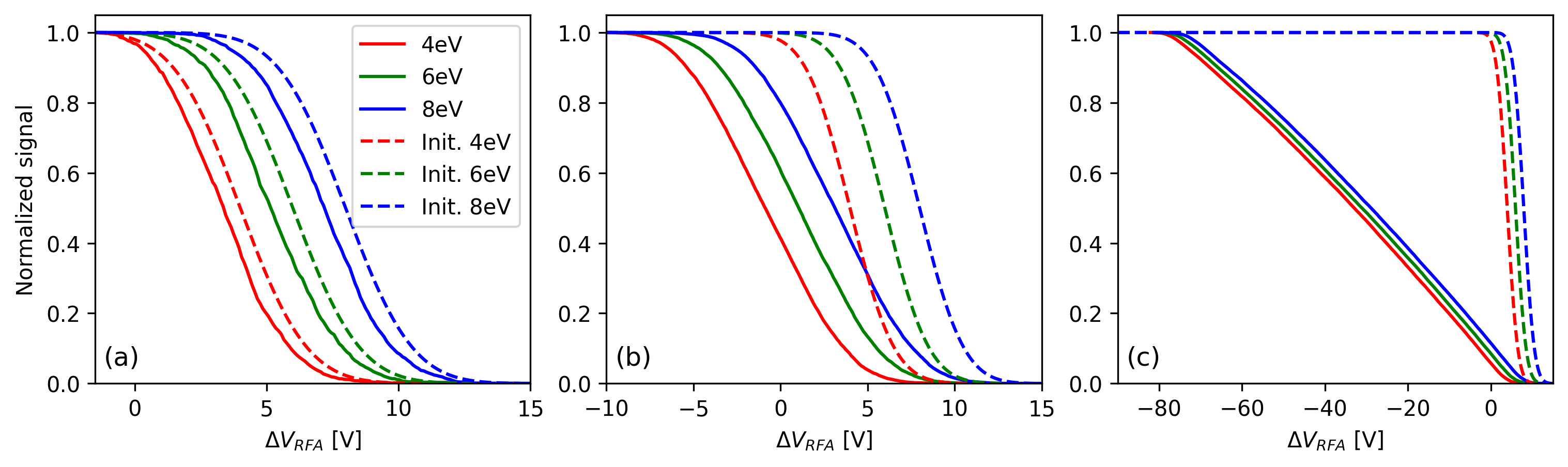}
    \caption{Simulated RFA IV-curves for different initial longitudinal energies of O$^+$ with (a) 2\% (b) 14\% and (c) 100\% fraction of the ions closest to the beam axis considered.}
    \label{fig:long energy effect}
\end{figure}

\paragraph{Longitudinal energy}
To study the effect of the plasma potential on the synthetic RFA IV-curves, ions were launched with 4, 6 and \SI{8}{\electronvolt} mean longitudinal energy with an energy spread of \SI{2}{\electronvolt}. The results are presented in figure~\ref{fig:long energy effect} for (a) 2\% (b) 14\% and (c) 100\% beam fractions closest to the axis. For the 2\% fraction, a very good match is obtained between the initial and final distributions and the pivot points are found at \SI{0.8}{\volt} lower than the mean energy of the ions. The same shift applies regardless of the initial energy. Again, this implies that the pivot point is a reasonable estimate of the plasma potential for tightly collimated beams. The relative error of the plasma potential values derived by considering the IV-curve pivot point becomes smaller at high plasma potential because the deviation remains constant at \SI{0.8}{\volt}. In the case of the 10\% fraction, the pivot point is \SI{3.5}{V} below the initial longitudinal energy, and the final distributions differ from the initial ones. These results imply that the plasma potential can be deduced from the RFA IV-curves taking into account a shift proportional to the collimation whereas the energy distributions are reliable only for tightly collimated beams. This is highlighted in table~\ref{tab:long energy} comparing different methods of deriving the plasma potential (referred as longitudinal energy here) from the RFA IV-curve. The pivot point underestimates the plasma potential up to 15\% when appropriate collimation is applied. When less severe collimation is applied, the most representative method is the linear fit.

\begin{table}[H]
    \begin{center}
    \begin{tabular}{|c|c c c|} 
    \hline
    Longitudinal energy (\SI{}{\electronvolt}) & 4 & 6 & 8 \\ [0.5ex] 
     \hline
    Pivot point 2\% fraction (\SI{}{\volt}) & 3.4 & 5.2 & 7.2 \\ 
    Pivot point 14\% fraction  (\SI{}{\volt}) & $-$0.9 & 1.0 & 3.1 \\
    Pivot point 100\% fraction  (\SI{}{\volt}) & $-$32.9  & $-$31.1 & $-$28.9  \\ 
    \hline
    Max dI/dV 2\% fraction (\SI{}{\volt}) & 4.1 & 4.5 & 7.3  \\ 
    Max dI/dV 14\% fraction  (\SI{}{\volt}) & $-$1.8 & 1.3 & 3.8 \\
    Max dI/dV 100\% fraction  (\SI{}{\volt}) & $-$3.0 & 0.1 & 1.5 \\ 
    \hline
    Zero cross 2\% fraction (\SI{}{\volt}) & 6.1 & 8.1 & 9.9  \\ 
    Zero cross 14\% fraction  (\SI{}{\volt}) & 4.3 & 6.2 & 8.2  \\
    Zero cross 100\% fraction  (\SI{}{\volt}) & 5.1 & 6.9 & 9.2   \\ 
    \hline    
    \end{tabular}
    \end{center}
    \caption{The values of the pivot point, maximum of dI/dV and IV-curve linear fit zero-crossing as a function of the imposed longitudinal energy of the beam, i.e.\ plasma potential.}    
    \label{tab:long energy}
\end{table}

\begin{figure}[H]
    \centering
    \includegraphics[width=\textwidth]{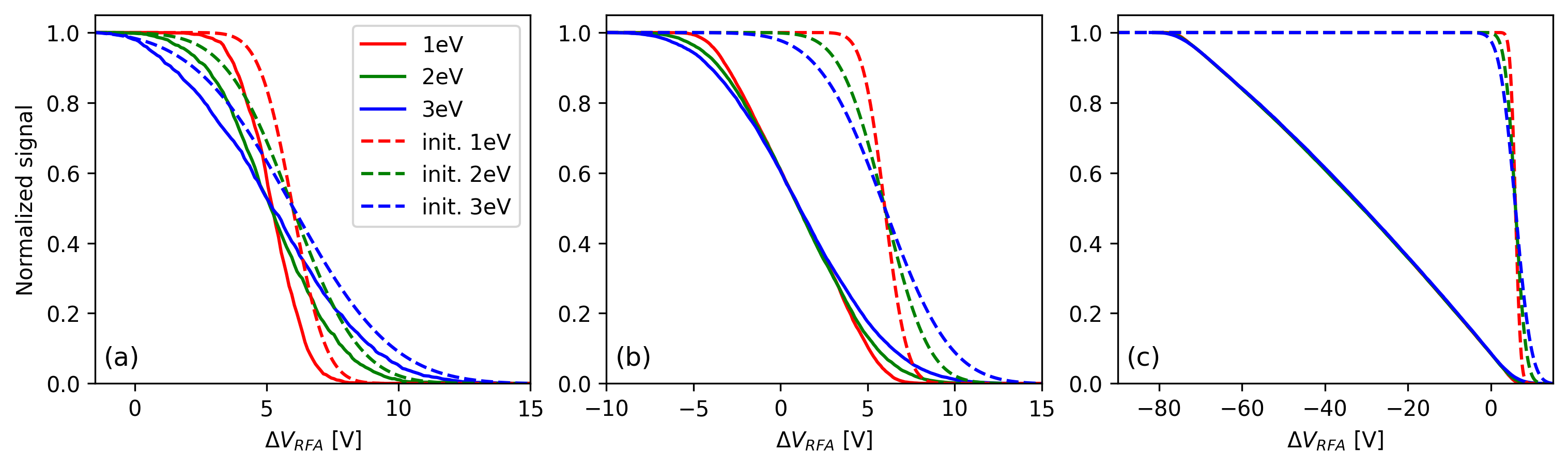}
    \caption{Simulated RFA IV-curves for different initial longitudinal energy spreads of O$^+$ with (a) 2\% (b) 14\% and (c) 100\% fraction of ions closest to the beam axis considered.}
    \label{fig:long energy spread effect}
\end{figure}

\paragraph{Initial energy spread of the ions}
Simulations were carried out varying either the initial longitudinal or transverse energy spread of the ions launched with a mean energy of \SI{6}{\electronvolt} corresponding to the plasma potential. The energy spread values used here were 1, 2 and \SI{3}{\electronvolt} either longitudinal or transverse, depicting the effect of the ion temperature and collisional sheath. The results in the two cases were virtually identical, so we present simulations only for the longitudinal energy spread. For the 2\% ions closest to the beam axis, see figure~\ref{fig:long energy spread effect}(a), the simulated curves are in good agreement with the initial distributions (dashed lines). With larger fractions, namely 14\% and 100\% as shown in figures~\ref{fig:long energy spread effect}(b) and \ref{fig:long energy spread effect}(c), the energy distributions differ significantly from the initial ones. The effect of the initial energy spread is observed as a changing curvature of the IV-curve tail near the maximum energy regardless of the collimation as argued in ref.~\cite{Tarvainen2005} where the curvature was used as a measure of the "effective ion temperature" of the beam. The data demonstrates that, with appropriate collimation the energy distribution caused by plasma effects can be accurately measured by the RFA regardless of the longitudinal energy spread.

\paragraph{Ion mass}

\begin{figure}[H]
     \centering
     \includegraphics[width=\textwidth]{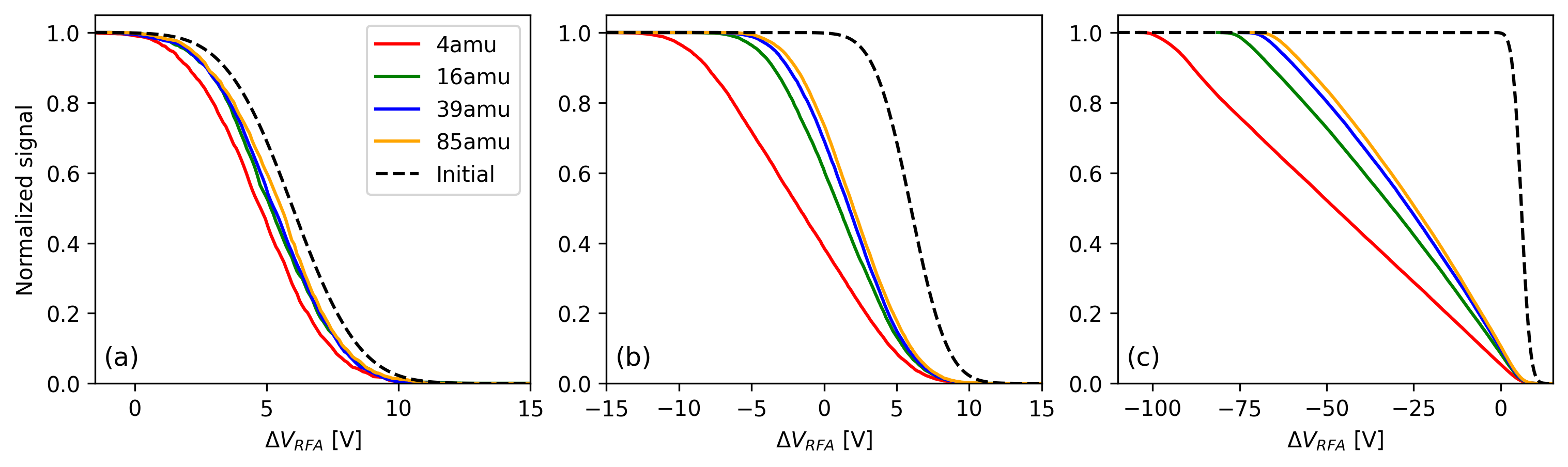}
    \caption{ Simulated RFA IV-curves for different ion masses with (a) 2\% (b) 14\% and (c) 100\% fraction of ions closest to the beam axis considered.}
    \label{fig:mass effect}
\end{figure}

To study the effect of the ion mass on the predicted RFA IV-curves, simulations were done for singly charged ions of 4, 16, 39 and 85 atomic mass units (the most abundant isotopes of He, O, K and Rb). The results are shown in figure~\ref{fig:mass effect}. The distributions are similar to the initial one when 2\% of the ions nearest to the axis are considered in (a). Again, with the 14\% and 100\% fractions the distributions differ from the initial distribution, the effect being pronounced for the lightest ion species due to the mass-dependent magnetic field induced rotation in the extraction and the corresponding longitudinal energy spread (see figure~\ref{fig:meffect} for comparison). 

\begin{figure}[H]
    \centering
    \includegraphics[width=\textwidth]{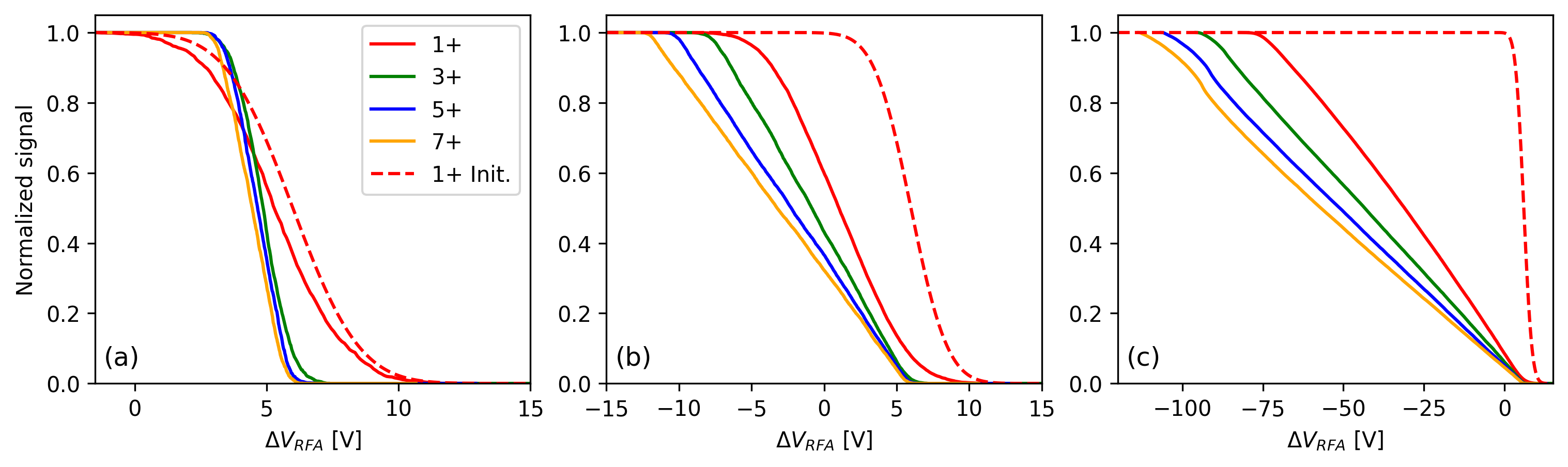}
    \caption{Simulated RFA IV-curves for different oxygen charge states with (a) 2\% (b) 14\% and (c) 100\% fraction of the ions closest to the beam axis considered.}
    \label{fig:charge state effect}
\end{figure}

\paragraph{Charge state}
Finally, the RFA IV-curves of oxygen ion charge states 1+, 3+, 5+ and 7+ were simulated. The initial ion energy was set to q*\SI{6}{\volt}, replicating the effect of the plasma potential, and the energy spread to \SI{2}{\electronvolt}. Figure~\ref{fig:charge state effect} shows the results for the 2\% (a), 14\% (b) and 100\% (c) beam fractions. The pivot point values are compared in table~\ref{tab:charge state effect}. There are a few points to be made. The shape of the IV-curve depends on the charge state as the effect of the ion temperature (initial energy spread), which was assumed to be the same for all charge states, is suppressed by the retarding potential with increasing ion charge. The pivot point method is valid only with severe collimation and even then it underestimates the plasma potential for high charge states. On the other hand, the linear fit method becomes more accurate with increasing charge state. 

\begin{table}[H]
    \begin{center}
    \begin{tabular}{|c|c c c c|} 
    \hline
    Charge state & 1+ & 3+ & 5+ & 7+ \\ [0.5ex] 
     \hline
     Pivot point 2\% fraction (\SI{}{\volt}) & 5.3 & 4.9 & 4.7 & 4.5 \\ 
    Pivot point 14\% fraction (\SI{}{\volt}) & 1.0 & $-$0.8 & $-$2.3 & $-$3.2 \\
    Pivot point 100\% fraction (\SI{}{\volt}) & $-$31.0 & $-$43.4 & $-$51.1 & $-$57.0 \\  
    \hline
    Max dI/dV 2\% fraction (\SI{}{\volt}) & 5.0 & 5.0 & 4.2 & 4.7 \\ 
    Max dI/dV 14\% fraction  (\SI{}{\volt}) & 2.0 & $-$3.3 & $-$9.4 & 3.2 \\
    Max dI/dV 100\% fraction  (\SI{}{\volt}) & $-$1.8 & $-$86.6 & $-$91.3 & $-$93.7 \\ 
    \hline
    Zero cross 2\% fraction (\SI{}{\volt}) & 8.3 & 6.1 & 5.9 & 5.7 \\ 
    Zero cross 14\% fraction  (\SI{}{\volt}) & 6.3 & 5.8 & 5.8 & 5.8 \\
    Zero cross 100\% fraction  (\SI{}{\volt}) & 6.9 & 6.4 & 5.5 & 5.2 \\ 
   
    \hline
    \end{tabular}
    \end{center}
    \caption{The values of the pivot point, maximum of dI/dV and IV-curve linear fit zero-crossing for different oxygen charge states.}    
    \label{tab:charge state effect}
\end{table}

Some key conclusions can be drawn from the synthetic RFA IV-curve simulations. A good collimation allows estimating the plasma potential with the pivot point method although the pivot point is found at a value slightly lower than the plasma potential. The accurate value of the plasma potential can be derived by applying an offset, which depend on the ion mass and charge state but is not sensitive to the source potential, extraction B-field (in normal operational range), or the initial energy spread (longitudinal or transverse). For example, in the typical CB operating conditions with \SI{20}{\kilo\volt} source potential and \SI{0.8}{\tesla} extraction B-field the offset value is \SI{1.2}{\volt} for He$^+$ and \SI{0.8}{\volt} for O$^+$. Simulations are required to deduce the exact offset for each ion species.
The shape of the IV-curves is affected by the initial energy spread, i.e. the ``effective temperature'' of the beam can be used as a coarse measure of the ion temperature.

\section{Experimental setup and results}

The motivation to study the longitudinal energy spread of the ion beams extracted from the LPSC CB stems from the desire to improve the understanding of the 1+ ion capture process into the ECRIS plasma. In particular, the goal is a comparison of the so-called $\Delta V$-curves of the high charge state ions reflecting the capture efficiency and the energy spread of the N+ ion beams extracted from the CB, possibly challenging the prevailing model for the ion capture based on slowing down in consecutive ion-ion collisions and requiring to match the velocity of the incident ions to the average velocity of the plasma ions~\cite{Geller,Tarvainen2022Capture}. For this purpose, and to measure the longitudinal energy spread due to the plasma characteristics (not due to the extraction conditions), only the ions closest to the beam axis are to be considered. In this section we describe the experimental setup and results corroborating the simulations presented above.

\subsection{CB ECRIS test bench at LPSC}

\begin{figure}[H]
    \centering
    \includegraphics[width=0.8\textwidth]{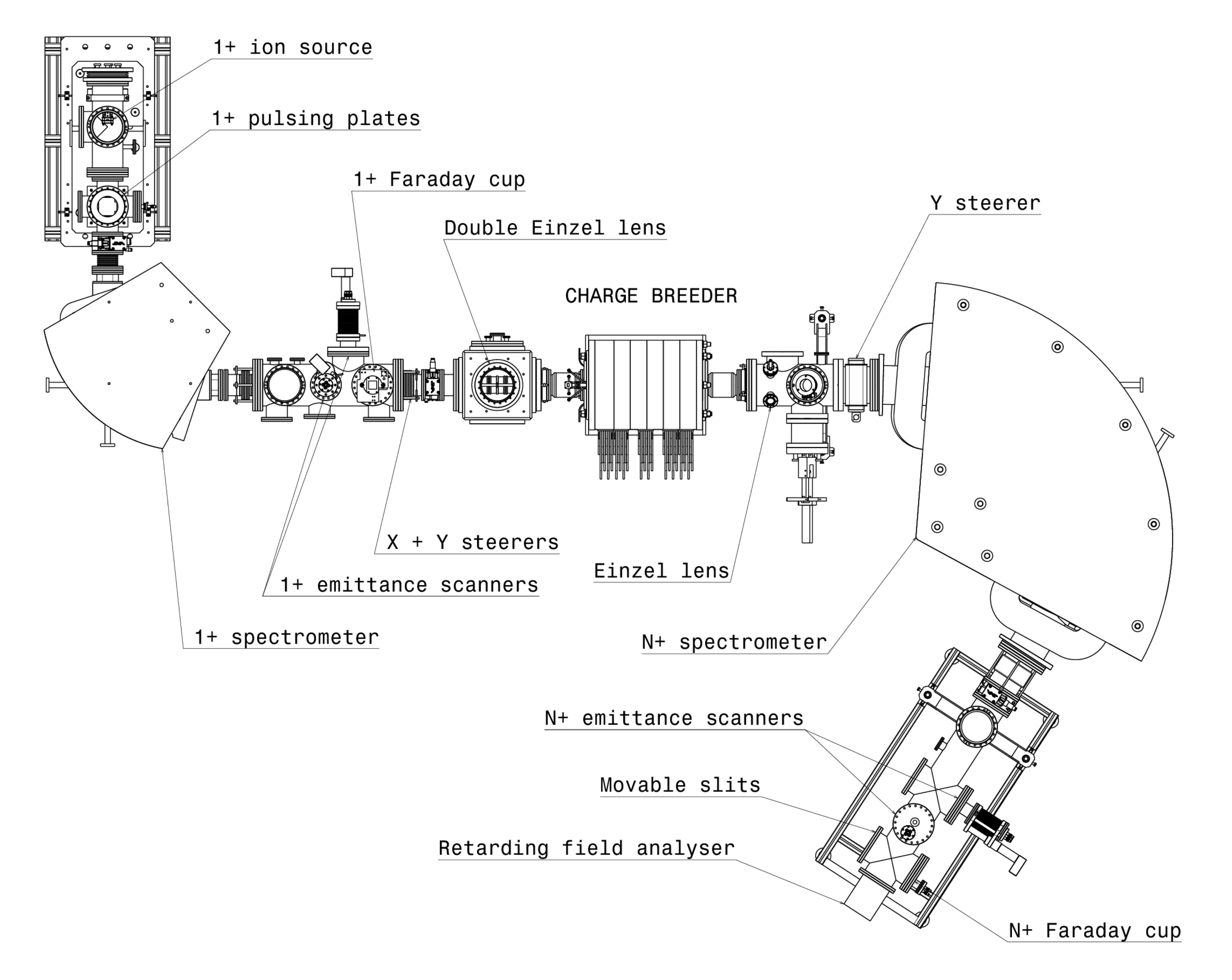}
    \caption{Schematic view of the 1+N+ test bench at LPSC with the main components labelled.}
    \label{fig:1+N+ test bench}
\end{figure}

LPSC has developed the PHOENIX type CB ECRIS since the early 2000s for applications in ISOL facilities~\cite{Maunoury2020,Ames2016,Galata2015}. This \SI{14.5}{\giga\hertz} ECRIS, described thoroughly in the literature ~\cite{Angot2020}, is equipped with 3 coils and a magnetic steel yoke to generate the asymmetric magnetic bottle. The injection and extraction coils drive the maxima of the axial magnetic field at injection B$_{\textrm{inj}}$ and extraction B$_{\textrm{ext}}$, respectively. The role of the middle coil is to (almost) independently adjust the minimum axial magnetic field B$_{\textrm{min}}$. The radial magnetic field is generated with a permanent magnet sextupole surrounding the plasma chamber. At LPSC, the 1+N+ test bench is dedicated for improving the PHOENIX-type CB performances, i.e. efficiency and charge breeding time. The "1+ beam line" is used for the injection of the 1+ beam into the CB. To test the 1+N+ method with alkali elements, a surface ionisation source is typically used as the 1+ source. The 1+ beam line is composed of a pulsed deflector to control the 1+ beam injection, a \SI{90}{\degree} dipole magnet, a diagnostics chamber and, finally, a double einzel lens to focus the beam at the CB injection port. At the CB extraction, there is an einzel lens adjacent to the puller electrode. Downstream, the N+ beam line continues with a \SI{120}{\degree} dipole magnet spectrometer and the N+ diagnostics chamber housing a Faraday cup and a retarding field analyser. The layout of the test bench is shown in figure~\ref{fig:1+N+ test bench}.

\subsection{Retarding field analyser} \label{ExperimentsImprovedRFA}
A modified version of the RFA developed at JYFL~\cite{JYFL_RFA}
 was installed at the end of the N+ beam line. The device was upgraded to withstand up to \SI{25}{\kilo\volt} potential from the initial \SI{10}{\kilo\volt} design for the CB diagnostics by scaling its dimensions. 
 
The original RFA is composed of a front plate electrode, a retarding grid with square holes \SI{1}{\milli\meter} apart with \SI{0.3}{\milli\meter} diameter wire and a collector plate. A large diameter collimating aperture was mounted upstream from the front plate in order to prevent stray ions reaching the collector around the front plate. As this RFA was conceived to estimate the maximum value of the ECR source plasma potential, it is not equipped with a repelling grid to suppress secondary electron because this application requires only a small variation of the grid potential around the $V_s$ value. The variation of the potential by some tens of volts is insignificant in comparison to the ion source potential of \SI{10}{\kilo\volt} or higher, thus resulting in the secondary electron emission from the collector being constant in the required voltage sweep (ion energy) range.

The potential difference $\Delta V_{RFA}$  between the CB plasma chamber ($V_s$) and the RFA grid was set using a tunable voltage supply, see figure~\ref{fig:JYFL RFA}. As $V_p$ is a small positive potential, setting $\Delta V_{RFA}$ at zero should allow all the ions to pass the retarding grid, and increasing $\Delta V_{RFA}$ to small positive value should induce their stopping. However, taking into account the prediction of the extraction simulations in previous sections, the energy distribution of the N+ ion beam is expected to extend to values lower than $qV_s$. Therefore, to allow the measurement of the longitudinal energy spread, a negative supply continuously delivering \SI{-60}{\volt} potential was set in series with the tunable positive power supply. Both these power supplies are floating on the ion source potential, which eliminates the ripple of the ion source bias power supply affecting the measured IV-curves.

\begin{figure}[H]
    \centering
    \includegraphics[width=0.5\textwidth]{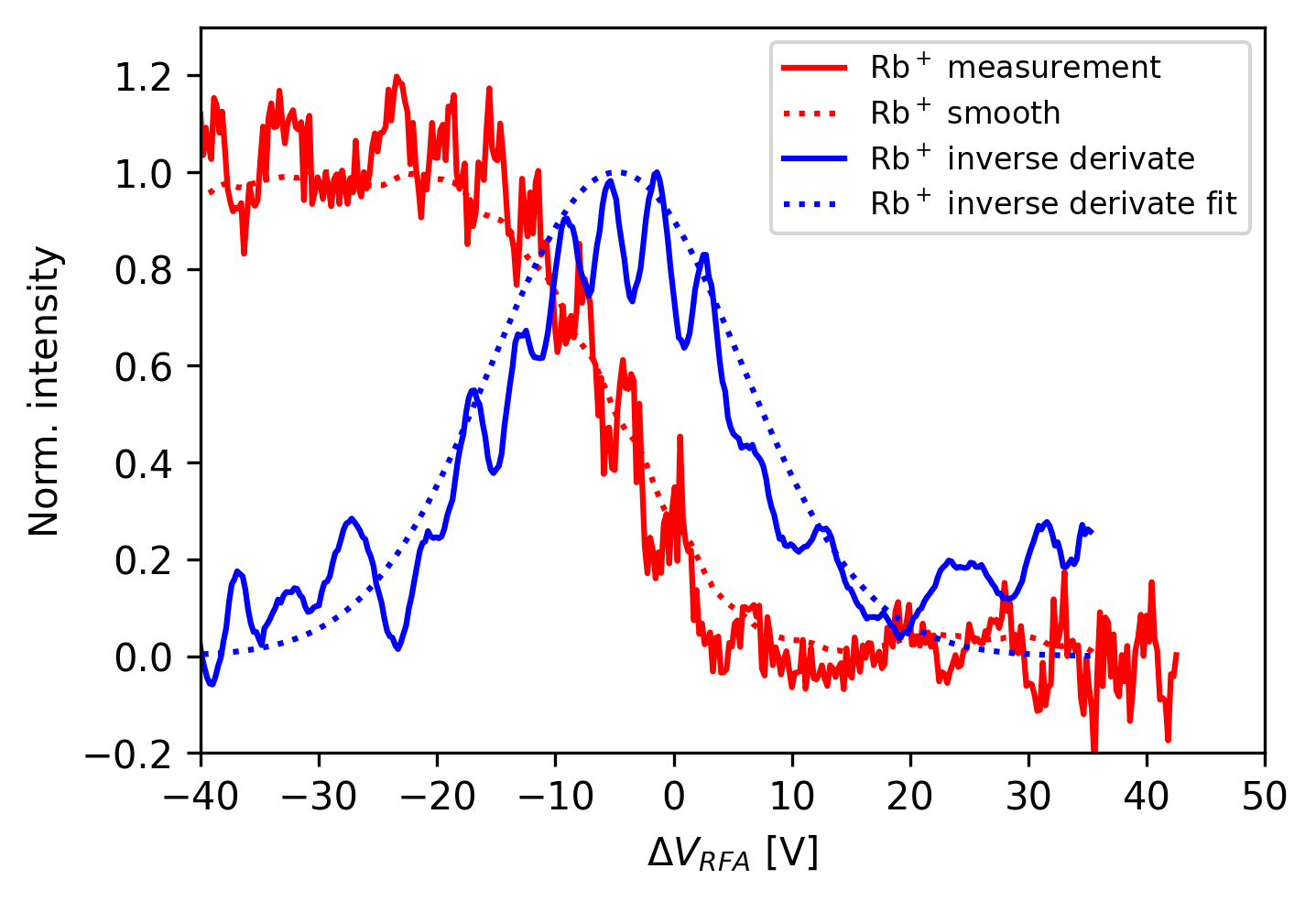}
    \caption{The Rb$^+$ RFA IV-curve (original RFA configuration), the inverse derivative of the IV-curve and the corresponding Gaussian fit.}
    \label{fig:RFA spectrum Rb+Rb+ 092021}
\end{figure}

The energy resolution of the RFA was estimated with a \SI{20}{\kilo\electronvolt} Rb$^+$ ion beam, produced with the 1+ surface ionisation source equipped with an Rb pellet manufactured by HeatWave Labs \cite{cathode}. Based on the temperature of the surface emitting the 1+ ions, their average energy when emitted can be estimated to be \SI{0.1}{\electronvolt}. It can therefore be argued that the initial energy spread of the 1+ beam is less than \SI{0.1}{\electronvolt}, which enables estimating the RFA energy resolution. The beam was transported through the test bench tuning the beam optics of the 1+ and N+ beam lines to maximise the beam intensity measured at the RFA collector. The CB plasma chamber was set to ground potential in this case to prevent electrostatic focusing effects whereas the CB coils were energised to focus the beam through the limiting apertures.  Figure~\ref{fig:RFA spectrum Rb+Rb+ 092021} shows the RFA IV-curve and the corresponding energy distribution with \SI{1}{\milli\meter} entrance aperture. The rolling average of the data (dashed curve) was then used to  obtain the RFA energy resolution (1$\sigma$) of \(\frac{\Delta E}{E}=\frac{\SI{10.5}{\electronvolt}}{\SI{20}{\kilo\electronvolt}}\).

Considering that the energy spread of the beams extracted from ECR ion sources should be resolved with \SI{1}{\electronvolt} rather than \SI{10}{\electronvolt} order of magnitude, the RFA resolution value was deemed too large. This motivated us to carry out simulations on the RFA itself with the goal of improving its energy resolution.
These simulations are presented in section \ref{RFADevelopment} of the appendices. Guided by these simulations, the retarding grid of the instrument was replaced by two superimposed and cross-aligned grids with smaller mesh size of \SI{0.44}{\milli\meter} and wire diameter of \SI{0.07}{\milli\meter}. The RFA entrance aperture diameter was set to \SI{1}{\milli\meter} to limit the off-axis ions being collected.

\begin{figure}[H]
    \centering
    \includegraphics[width=0.5\textwidth]{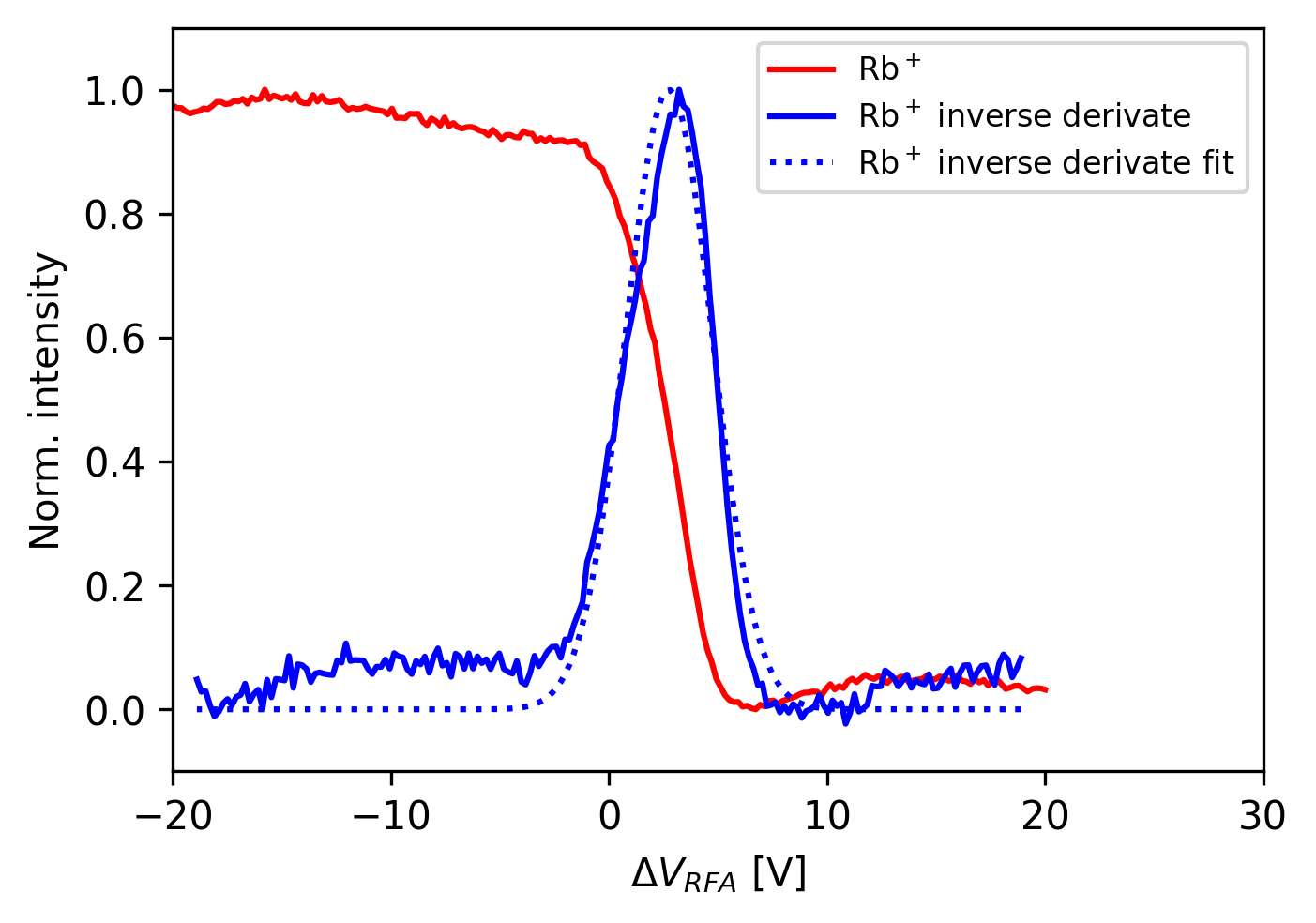}
    \caption{The Rb$^+$ RFA IV-curve (modified RFA configuration), the inverse derivative of the IV-curve and the corresponding Gaussian fit.}
    \label{fig:RFA spectrum Rb+ new}
\end{figure}

The RFA resolution measurement was repeated with the new configuration using the Rb$^+$ beam produced with the 1+ source. Figure~\ref{fig:RFA spectrum Rb+ new} shows the $Rb^+$ RFA IV-curve together with the Gaussian fit of the inverse derivative. The 1$\sigma$ energy resolution is \(\frac{\Delta E}{E}=\frac{\SI{2.0}{\electronvolt}}{\SI{20}{\kilo\electronvolt}}\) which represents an improvement by a factor of five with respect to the original configuration. Although the resolution does not exactly match the simulation results, the improvement factor is consistent with the prediction.

\begin{figure}[H]
     \centering
     \includegraphics[width=\textwidth]{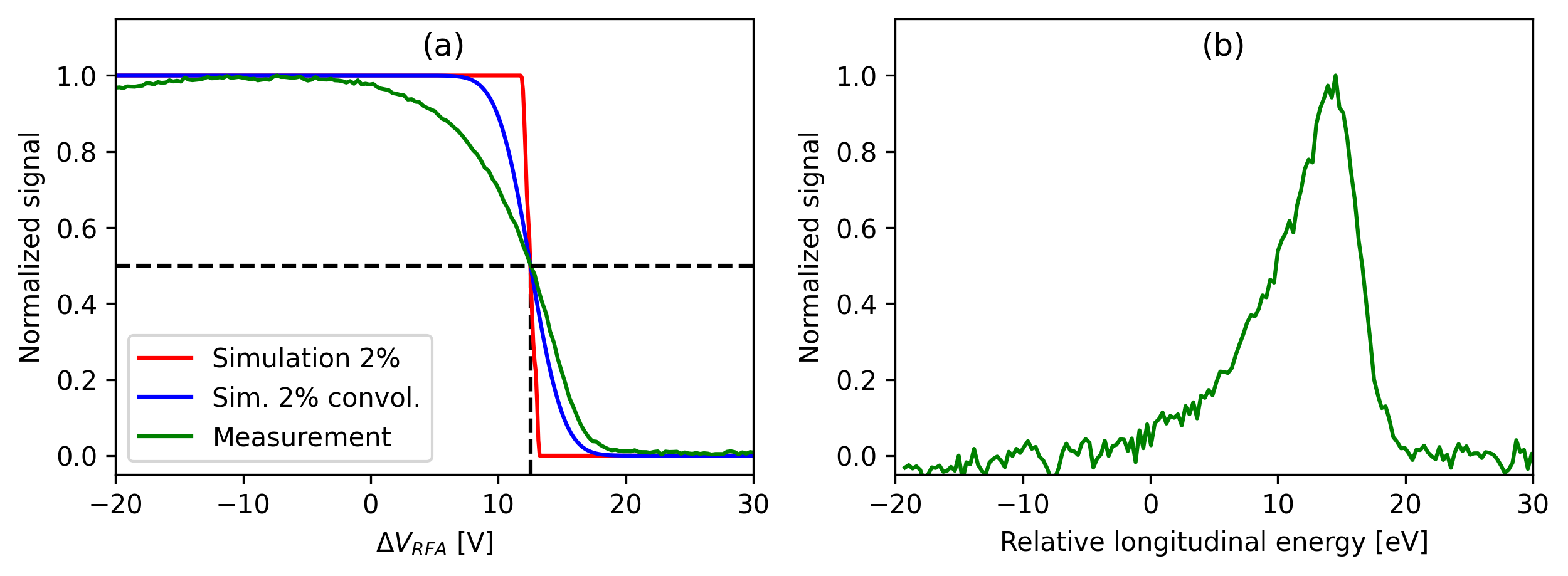}
    \caption{(a) O$^+$ RFA IV-curve measured with \SI{1}{\milli\meter} entrance aperture, simulation with 2\% beam fraction and the convoluted IV-curve taking into account the RFA energy resolution, and (b) the O$^+$ energy distribution.}
    \label{fig:RFA spectrum convol new}
\end{figure}

After estimating the energy resolution of the modified RFA, the O$^+$ RFA IV-curve was measured with the CB at \SI{20}{\kilo\volt}, \SI{550}{\watt} of microwave power and Helium as support gas. It is worth noting that in this case Oxygen was present as a contaminant in the CB plasma. The coil currents were set to generate an axial B field profile with B$_{\textrm{{inj}}}$=\SI{1.52}{\tesla}, B$_{\textrm{{min}}}$=\SI{0.40}{\tesla} and B$_{\textrm{{ext}}}$=\SI{0.85}{\tesla}. Figure~\ref{fig:RFA spectrum convol new}(a) shows the measured O$^+$ RFA IV-curve (green) with the simulation accounting for 2\% of the ions without any initial energy spread (red)\footnote{The 2\% ion fraction corresponds to experimentally found beam intensity fraction of the RFA and Faraday cup measurement}, and the convoluted 2\% simulation RFA spectrum (blue) taking into account the RFA resolution. The synthetic IV-curves of O$^+$, simulated with \SI{0}{\electronvolt} longitudinal energy, were shifted horizontally by +\SI{13.3}{\volt} to coincide with the pivot point of the measured IV-curve, which is at $\Delta V_{RFRA}$ of \SI{12.6}{\volt}. The plasma potential is \SI{13.3}{\volt} in this case.

\begin{figure}[H]
    \centering
    \includegraphics[width=\textwidth]{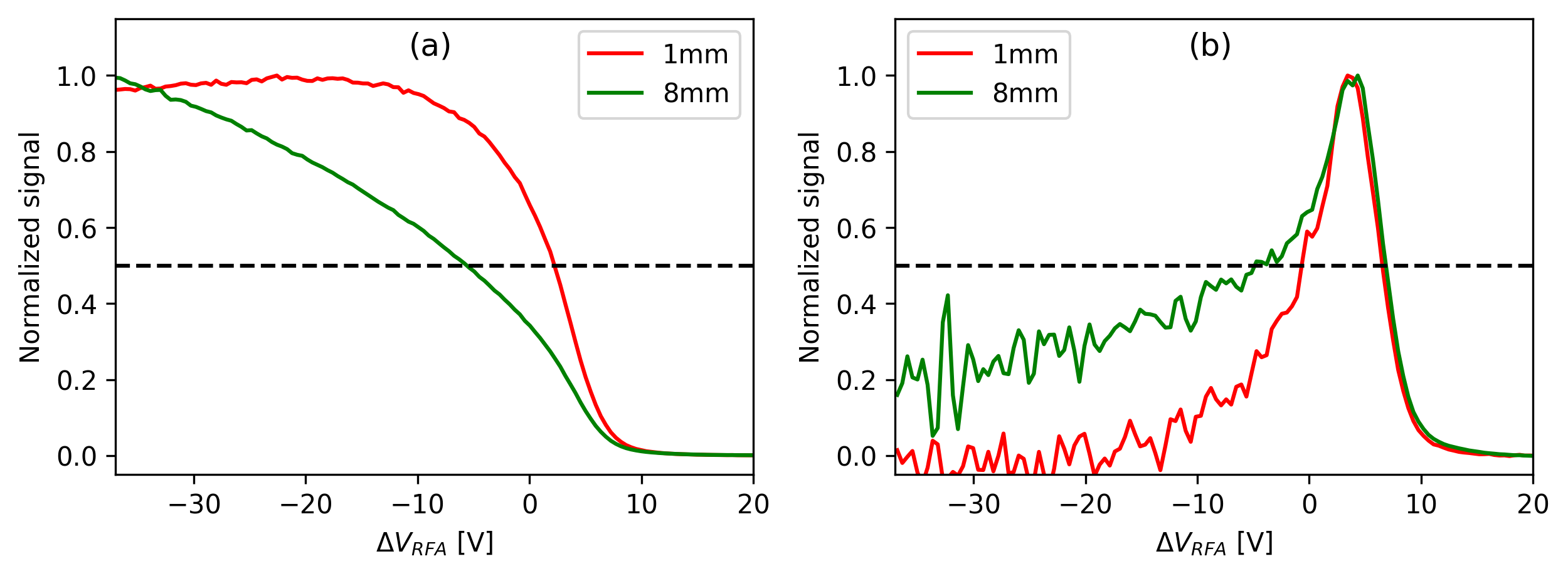}
    \caption{(a) O$^+$ RFA IV-curve with 2\% and 14\% of the incident beam entering the RFA with \SI{1}{\milli\meter} and \SI{8}{\milli\meter} entrance apertures, and (b) the corresponding longitudinal energy distributions}.
    \label{fig:RFA spectrum comp 1 8}
\end{figure}

The much improved resolution of the RFA allows comparing the energy spread with varying entrance collimator sizes, i.e. fractions of the incident beam. To achieve this, the RFA entrance hole was enlarged from \SI{1}{\milli\meter} to \SI{8}{\milli\meter}. The corresponding beam fractions, $\sim$2\% and $\sim$14\%, were measured with a Faraday cup using a similar collimation. The CB was again operated at \SI{20}{\kilo\volt} with He as support gas and an axial B-field profile with B$_{\textrm{{inj}}}$=\SI{1.50}{\tesla}, B$_{\textrm{{min}}}$=\SI{0.36}{\tesla} and B$_{\textrm{{ext}}}$=\SI{0.83}{\tesla}. The RFA spectrum of O$^+$, extracted from the CB as a contaminant, was recorded with \SI{8}{\milli\meter} RFA entrance aperture with and without an additional \SI{1}{\milli\meter} movable collimator, at \SI{280}{\watt} microwave. The corresponding RFA IV-curves and longitudinal  energy distributions are shown in figure~\ref{fig:RFA spectrum comp 1 8}.

Two conclusions are drawn: (i) The improved RFA with heavy collimation allows to estimate the plasma potential of the CB and the effect of various ion source tuning parameters on it; (ii) The energy spread of the ion beam depends on the fraction of ions measured, i.e.\ the experiment is in good agreement with the simulations suggesting that the longitudinal energy spread of the beam is dominated by the electrostatic focusing effect (and to some degree magnetic field induced rotation) whereas plasma effects are visible only for the ions extracted near the optical axis.

\section{Parametric RFA measurements on the CB ECRIS}

In the experimental part of this work we studied the effect of ion species and ion source potential on the recorded IV-curves to exclude them affecting the interpretation of parametric sweep results. In the parametric sweeps we varied the microwave power and CB magnetic field to detect their effect on the RFA IV-curves and plasma potential. The measured pivot point values were used for estimating the plasma potential, correcting the systematic error predicted by the simulations (see section \ref{RFA_simulations}) .

\subsection{Ion species and charge state}
\begin{figure}[H]
    \centering
    \includegraphics[width=\textwidth]{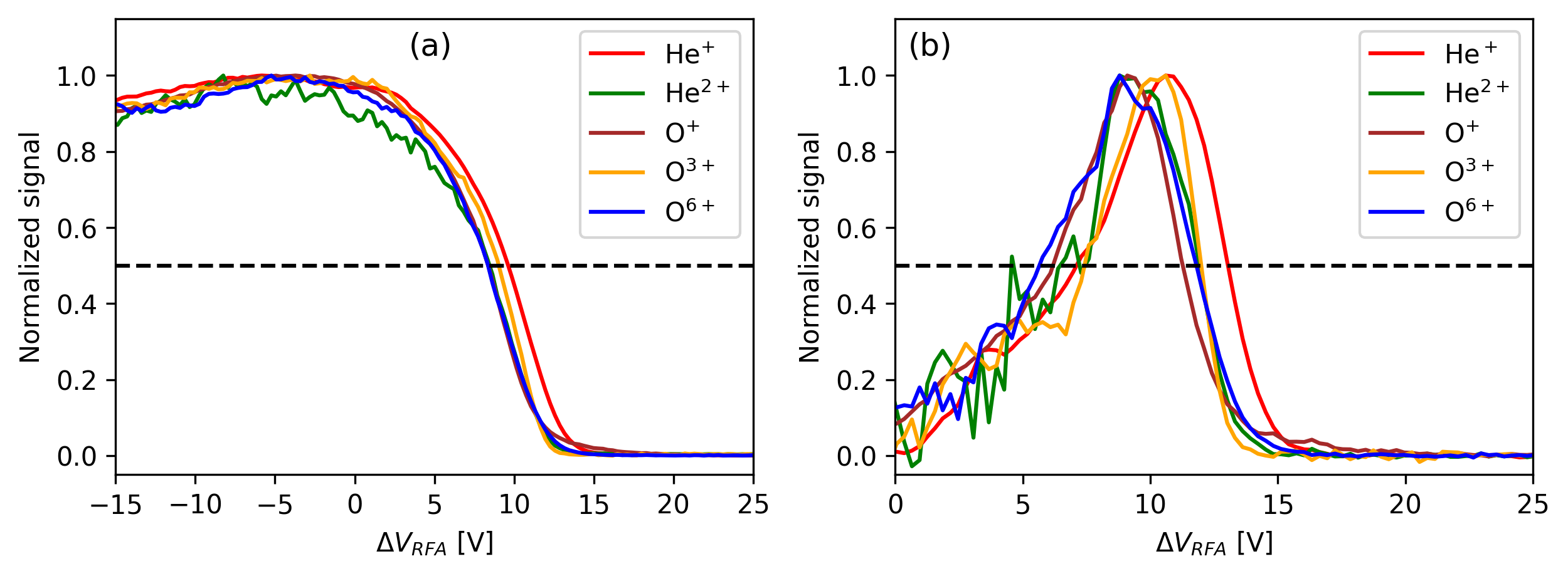}
    \caption{(a) The RFA IV-curves and (b) corresponding energy distributions of different ion species extracted from the same ECRIS plasma.}
    \label{fig:exp_results species}
\end{figure}

The RFA IV-curves of different charge states of helium and oxygen ions were measured first. The source was operated with He as support gas and \SI{300}{\watt} microwave power. The source potential was set to \SI{20}{\kilo\volt}. The oxygen ions are present as contaminants (impurities) in the He plasma. The CB coil currents were set to generate a magnetic field with B$_{\textrm{inj}}$, B$_{\textrm{min}}$ and B$_{\textrm{ext}}$ of \SI{1.53}{\tesla}, \SI{0.43}{\tesla} and \SI{0.83}{\tesla}, respectively. Figure~\ref{fig:exp_results species} presents the RFA IV-curves and their derivatives for the measured ion species. Table~\ref{tab:tab exp_species} summarises the pivot point values together with the FWHM of the energy distributions (derivative curves). Since all ions are extracted from the same plasma, the contribution of the plasma potential on their longitudinal energy can be expected to be the same. This is corroborated by the IV-curves for different ion species and charge states. The estimated plasma potential is 9--\SI{13}{\volt} with no apparent charge state dependence. It is therefore concluded that the RFA technique (and the improved device) is suitable for the measurement of the plasma potential with an uncertainty of $\pm$20\%.

\begin{table}[H]
    \begin{center}
    \begin{tabular}{|c|c c c c c|} 
     \hline
     Species & He$^+$ & He$^{2+}$ & O$^+$ & O$^{3+}$ & O$^{6+}$ \\ [0.5ex] 
     \hline
     Pivot point (\SI{}{\volt}) & 9.6 & 8.5 & 8.4 & 9.0 & 8.4 \\ 
     \hline
     Plasma potential (\SI{}{\volt}) & 10.8 & 13.2 & 9.1 & 10.1 & 9.8 \\
          \hline
     FWHM (\SI{}{\volt}) & 5.9 & 7.4 & 5.1 & 4.6 & 6.2   \\
     \hline
    \end{tabular}
    \end{center}
    \caption{The pivot point, plasma potential and FWHM of the energy distribution for different ion species.}
    \label{tab:tab exp_species}
\end{table}

\subsection{Source potential}
\begin{figure}[H]
    \centering
    \includegraphics[width=\textwidth]{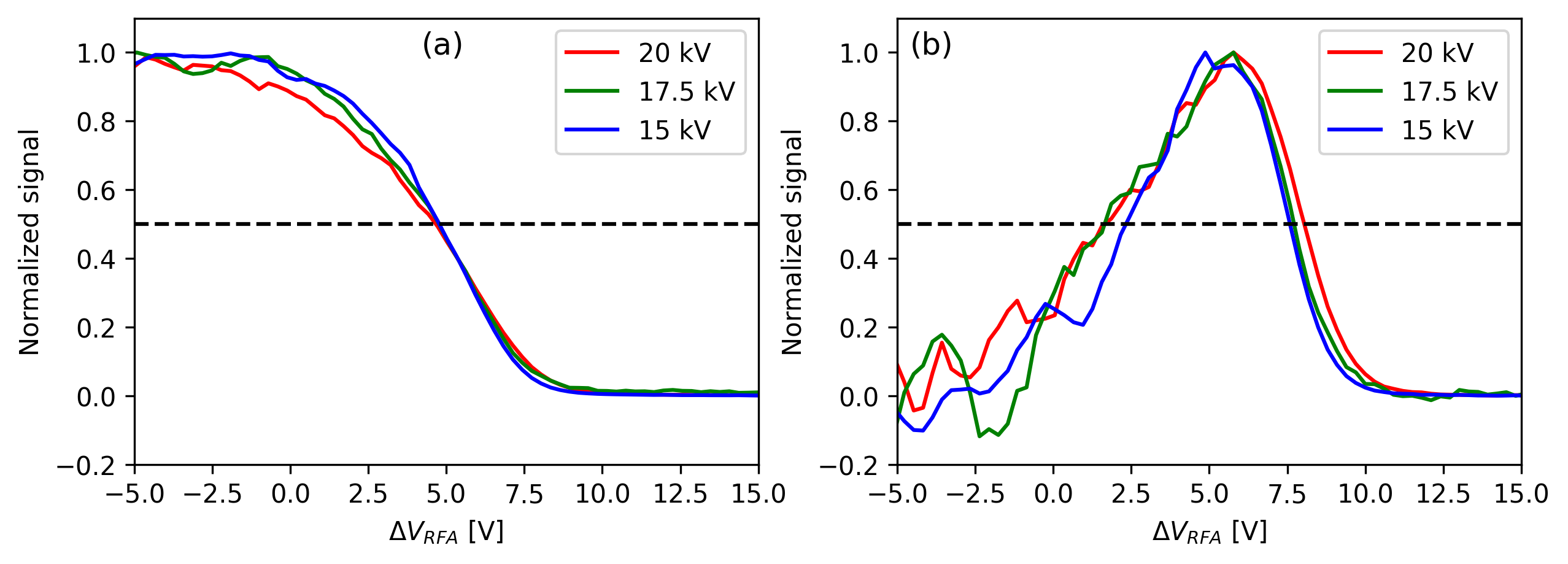}
    \caption{(a) The RFA IV-curves and (b) corresponding energy distributions of He$^+$ ions with ion source potentials of \SI{15}{\kilo\volt}, \SI{17.5}{\kilo\volt} and \SI{20}{\kilo\volt}.}
    \label{fig:exp_results hv}
\end{figure}

The effect of the source potential on the He$^+$ RFA IV-curves was studied at 15, 17.5 and \SI{20}{\kilo\volt}. Helium was used as the plasma support gas and the microwave power was set to \SI{223}{\watt}. The B$_{\textrm{inj}}$, B$_{\textrm{min}}$ and B$_{\textrm{ext}}$ values were \SI{1.53}{\tesla}, \SI{0.43}{\tesla} and \SI{0.83}{\tesla}, respectively. The He$^+$ RFA IV-curves and the corresponding energy distributions are plotted in figure~\ref{fig:exp_results hv}. The pivot points, estimated plasma potentials and FWHM of the energy distributions are summarised in table \ref{tab:tab exp_HT}. There is no difference between the IV-curves, so it is concluded that the ion source potential does not affect the plasma potential or the detection method at least with voltages near the nominal extraction potential. In this case the plasma potential is considerably lower than in figure~\ref{fig:exp_results species}, which is most likely due to the lower microwave power as discussed in detail hereafter.

\begin{table}[H]
    \begin{center}
    \begin{tabular}{|c|c c c|} 
     \hline
     $V_s$ (\SI{}{\kilo\volt}) & 15 & 17.5 & 20  \\ [0.5ex] 
     \hline
     Pivot point (\SI{}{\volt}) & 4.8 & 4.8 & 4.7\\ 
     \hline
     Plasma potential (\SI{}{\volt}) & 6.0 & 6.0 & 5.9  \\
          \hline
     FWHM (\SI{}{\volt}) & 5.3 & 6.1 & 6.4   \\
     \hline
    \end{tabular}
    \end{center}
    \caption{The pivot point, plasma potential and FWHM of the energy distribution of He$^+$ ions with different ion source potentials.}
    \label{tab:tab exp_HT}
\end{table}

\subsection{Microwave power}
\begin{figure}[H]
    \centering
    \includegraphics[width=\textwidth]{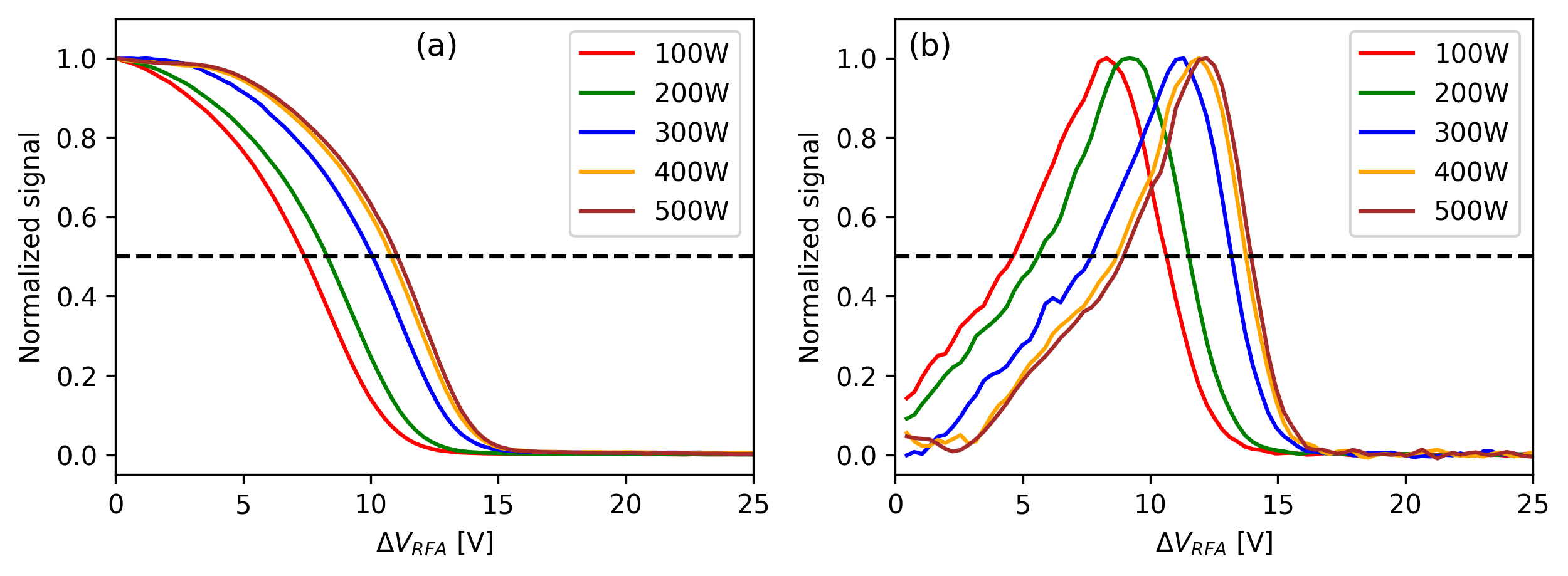}
    \caption{(a) He$^+$ RFA IV-curves and (b) the derivatives of the IV-curves as a function of microwave power.}
    \label{fig:exp_results hfpower}
\end{figure}

To study the effect of the injected microwave power on the RFA IV-curves, the CB was operated at \SI{20}{\kilo\volt} with He as support gas. The coil currents were set to produce an axial magnetic field with B$_{\textrm{{inj}}}$=\SI{1.53}{\tesla}, B$_{\textrm{{min}}}$=\SI{0.43}{\tesla} and B$_{\textrm{{ext}}}$= \SI{0.83}{\tesla}. Figure~\ref{fig:exp_results hfpower} presents the He$^+$ RFA IV-curves for microwave powers ranging from \SI{100}{\watt} to \SI{500}{\watt}. The pivot points and plasma potentials together with the FWHM-value of the distribution are summarised in table \ref{tab:tab MW power}.

\begin{table}[H]
    \begin{center}
    \begin{tabular}{|c|c c c c c|} 
    \hline
    Microwave power (\SI{}{\watt}) & 100 & 200 & 300 & 400 & 500 \\ [0.5ex] 
     \hline
     Pivot point (\SI{}{\volt}) & 7.4 & 8.3 & 10.1 & 10.8 & 11.1 \\ 
    \hline
    Plasma potential (\SI{}{\volt}) & 8.6 & 9.5 & 11.3 & 12.0 & 12.3 \\ 
    \hline
    FWHM (\SI{}{\volt}) & 6.0 & 5.9 & 5.5 & 5.1 & 5.0 \\ 
    \hline
    \end{tabular}
    \end{center}
    \caption{The pivot point, plasma potential and FWHM of the energy distribution of He$^+$ as a function of the microwave power.}    
    \label{tab:tab MW power}
\end{table}

The effect of the increasing microwave power can clearly be seen on the He$^+$ RFA spectrum, which validates the ability of the instrument to be used for studying parametric dependencies of the CB ECRIS. Increasing the microwave power from \SI{100}{\watt} to \SI{500}{\watt} results in approximately 50\% increase of the plasma potential from \SI{8.6}{\volt} to \SI{12.3}{\volt}. At the same time, the FWHM of the distribution decreases by less than 20\%. The shift of the IV-curve, i.e.\ increase of the beam longitudinal energy (plasma potential) with the increase of the microwave power, is consistent with earlier experiments on conventional ECR ion source reported in the literature~\cite{JYFL_RFA}. The most likely cause of the higher plasma potential is increased plasma density, which translates to increased charge breeding efficiency at high microwave powers~\cite{Emilie}.

\subsection{Extraction magnetic field}
\begin{figure}[H]
    \centering
    \includegraphics[width=\textwidth]{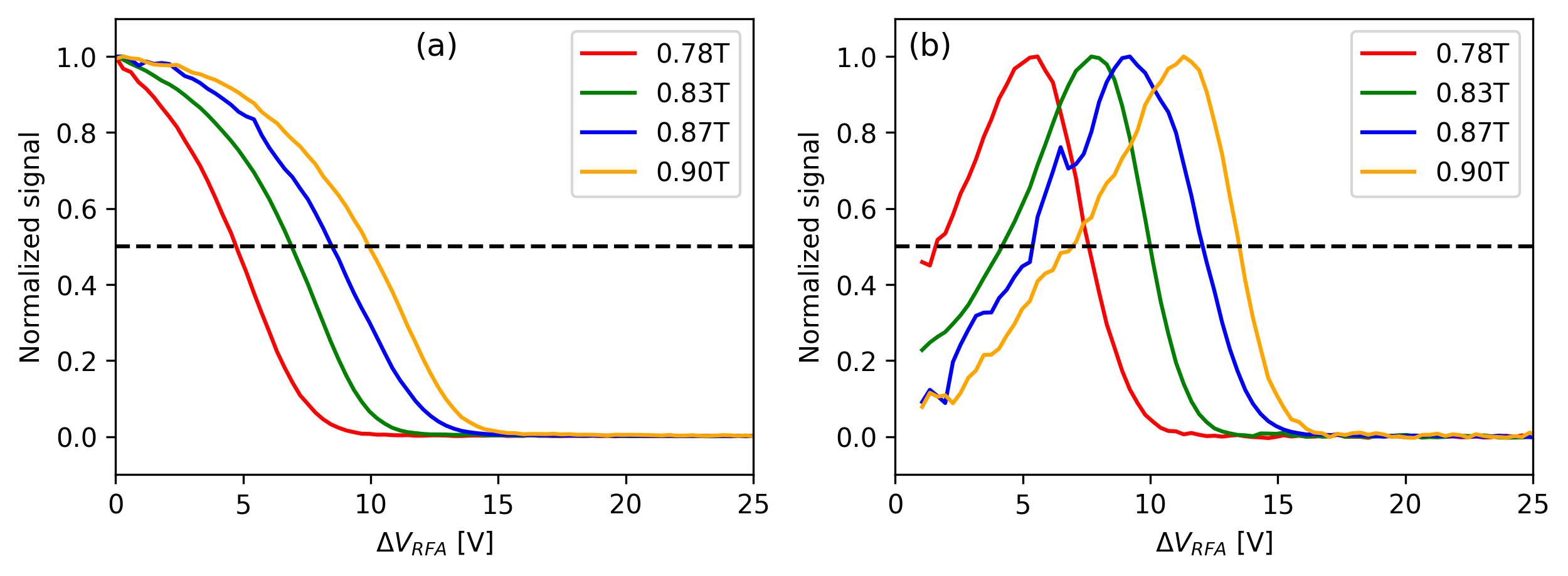}
    \caption{(a) He$^+$ RFA IV-curves and (b) the derivatives of the IV-curves as a function of the extraction magnetic field.}
    \label{fig:exp_ext coil current}
\end{figure}

Next, the He$^+$ RFA IV-curves were measured as a function of the extraction coil current, which changed the extraction magnetic field (maximum) from \SI{0.78}{\tesla} (\SI{592}{\ampere}) to \SI{0.90}{\tesla} (\SI{742}{\ampere}), keeping the injection and minimum-B fields virtually unchanged at \SI{1.53}{\tesla} and \SI{0.42}{}--\SI{0.44}{\tesla}. The CB was operated at \SI{20}{\kilo\volt} with He as support gas and \SI{300}{\watt} of microwave power. Figure~\ref{fig:exp_ext coil current} shows the recorded He$^+$ RFA IV-curves. The corresponding pivot points and plasma potentials together with the FWHM-value of the distribution are summarised in table \ref{tab:tab MW power}.

\begin{table}[H]
    \begin{center}
    \begin{tabular}{|c|c c c c|} 
    \hline
    Extraction coil current (\SI{}{\ampere}) & 592 & 625 & 712 & 742 \\
    Extraction field (\SI{}{\tesla}) & 0.78 & 0.83 & 0.87 & 0.90 \\ [0.5ex]
     \hline
     Pivot point (\SI{}{\volt}) & 4.7 & 6.9 & 8.5 & 9.9 \\ 
    \hline
    Plasma potential (\SI{}{\volt}) & 5.9 & 8.1 & 9.7 & 11.1 \\ 
    \hline
    FWHM (\SI{}{\volt}) & 6.0 & 5.8 & 6.7 & 6.5 \\ 
    \hline
    \end{tabular}
    \end{center}
    \caption{Characteristic values derived from the IV-curves and the FWHM of the distribution as a function of the extraction magnetic field.}    
    \label{tab:tab MW power}
\end{table}

Since the fraction of the beam entering the RFA is only 2\% (nearest to the axis), the effect of the extraction field on the longitudinal energy spread of the detected ions can be expected to be very small. Hence, it is concluded that the evident shift of the distribution with the extraction field strength is due to an increase of the plasma potential. The change of the plasma potential implies an improved ion confinement or a reduced (cold) electron confinement, as a higher ambipolar potential is required to compensate the loss rates of negative and positive charges. We note that the extraction field is the weakest magnetic mirror point of the CB ECRIS, so it can be expected to control the hot electron losses (see e.g. ref.~\cite{Toivanen} for further discussion) and global plasma confinement, thus, affecting the plasma potential.

\subsection{Injection magnetic field}

\begin{figure}[H]
    \centering
    \includegraphics[width=\textwidth]{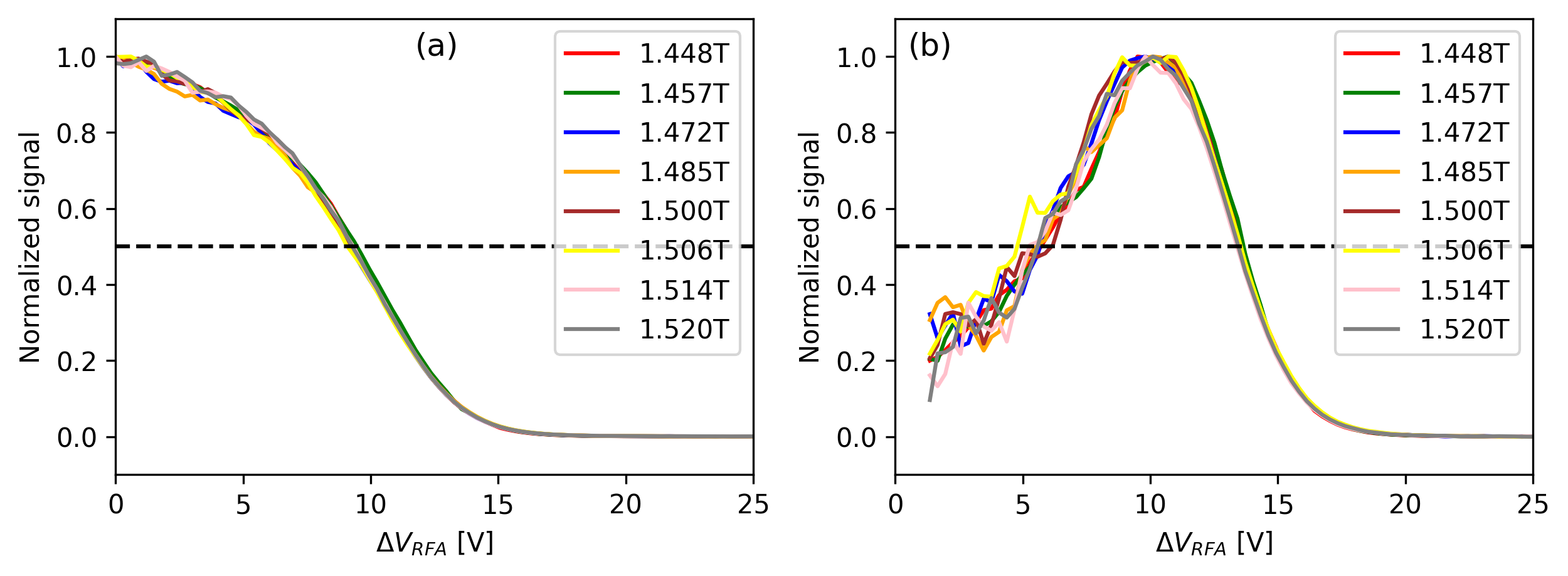}
    \caption{(a) He$^{2+}$ RFA IV-curves and (b) the derivatives of the IV-curves as a function of the injection magnetic field.}
    \label{fig:exp_results Binj}
\end{figure}

The effect of the (maximum) magnetic field at the injection, B$_{\textrm{inj}}$, on the RFA IV-curves was studied with He$^{2+}$ beam. We note that the selection of the charge state does not change the result as shown earlier. The minimum and extraction fields, B$_{\textrm{min}}$ and B$_{\textrm{ext}}$ were kept constant at respectively \SI{0.42}{\tesla} and \SI{0.81}{\tesla} by adjusting the coil currents based on simulations with \textsc{Radia3D} software. For the injection field sweep the source was operated with helium as support gas, with ion source potential of \SI{20}{\kilo\volt} and microwave power of \SI{260}{\watt}. Figure~\ref{fig:exp_results Binj} shows the obtained He$^{2+}$ RFA IV-curves (a) and energy distributions (b) when B$_{\textrm{inj}}$ was gradually changed from \SI{1.448}{\tesla} to \SI{1.520}{\tesla}. There is virtually no change at all with the plasma potential, the pivot point being in the range 9.2--\SI{9.5}{\volt}. Thus, we conclude that adjusting the stronger axial mirror field has very little effect on the plasma dynamics.

\subsection{Minimum magnetic field}

\begin{figure}[H]
    \centering
    \includegraphics[width=\textwidth]{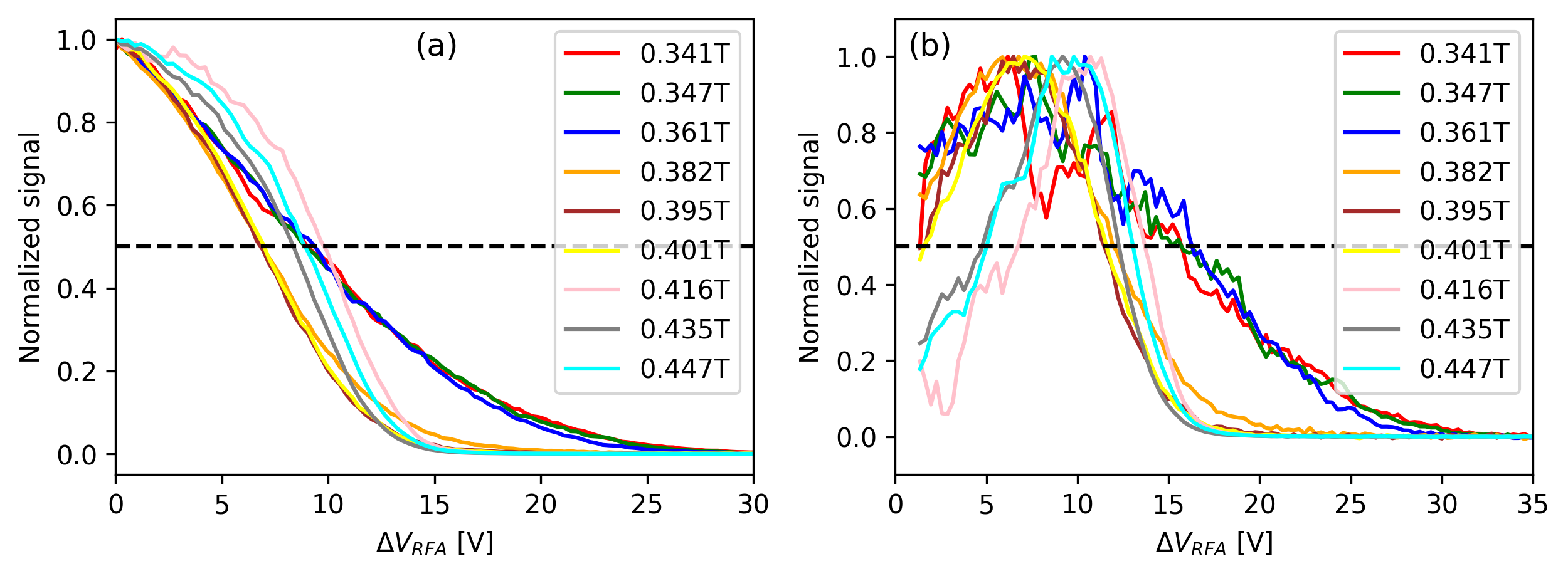}
    \caption{(a) He$^{2+}$ RFA IV-curves and (b) the derivatives of the IV-curves as a function of the minimum magnetic field.}
    \label{fig:exp_results Bmed}
\end{figure}

As a last parametric sweep we studied the effect of the minimum magnetic field, B$_{\textrm{min}}$, on the RFA IV-curves with He$^{2+}$ beam. In this case the injection and extraction fields were set at respectively \SI{1.50}{\tesla} and \SI{0.81}{\tesla}. The other ion source parameters were the same as in the injection field sweep. The RFA IV-curves and corresponding energy distributions are presented in figure~\ref{fig:exp_results Bmed}. In this case we observed a step-change in the energy distribution when the B$_{\textrm{min}}/$B$_{\textrm{ECR}}$-value exceeds 0.8, which is the expected optimum for high charge state ion production according to the semi-empirical ECRIS scaling laws~\cite{Hitz}. Below this threshold the energy spread of the He$^{2+}$ beam is larger. Interestingly, the beam current collected by the RFA increased by 80\% at B$_{\textrm{min}}$-values above this threshold, which could indicate a change in the plasma stability regime. It has been shown that the charge breeding efficiency of high charge state ions increases monotonically with B$_{\textrm{min}}$ until an instability threshold is reached around B$_{\textrm{min}}/$B$_{\textrm{ECR}}$ of 0.8~\cite{CB_instability}. The plasma potential values derived from the IV-curves are 12.8--\SI{15.8}{\volt} without a systematic dependence on B$_{\textrm{min}}$. The B$_{\textrm{min}}$-value has been shown to be the most critical field parameter affecting the hot electron tail of the electron energy distribution and plasma stability~\cite{Benitez, Izotov, Tarvainen_instability}. The result presented here suggests that the effect of B$_{\textrm{min}}$ on the cold electron population, which dictates the formation of the ambipolar plasma potential, is not significant. 

\section{Discussion}

We have presented a comprehensive study of factors affecting the longitudinal energy spread of ion beams extracted from an ECRIS (charge breeder). Both, simulations and experiments with the RFA support the following conclusions:

\begin{itemize}
    \item The longitudinal and transverse energy (momentum) spread of the beams is affected by the electrostatic focusing effects in the extraction, i.e.\ extraction geometry and plasma beam boundary, to the extent that it dominates over the magnetic field induced rotation of the beam or the effect of plasma potential and ion temperature.
    \item The electrostatic focusing effect on the longitudinal energy spread is more prominent than the temporal variation of the beam energy caused by the high-voltage ripple. The magnitude of the ripple with typical high-voltage power supplies is comparable to the energy spread due to the magnetic field induced rotation of the beam or the effect of plasma potential and ion temperature. We emphasize that the temporal variation of the ion source potential affects the time-averaged energy spread of the beam, not the momentary energy spread (modelled by the simulations in this paper) or the RFA measurements where the ripple is eliminated by floating the RFA grid power supply as the time-of-flight between the CB-ECRIS and the RFA is shorter than the period of the high-voltage ripple.
    \item Measurement of the ion beam energy spread, relevant for the downstream accelerator, requires collecting all ions. On the contrary, studying the effect of plasma properties (plasma potential and ion temperature) on the longitudinal energy spread requires heavy collimation of the beam accepting only ions near the symmetry axis of the beam.
    \item None of the normally applied methods for determining the plasma potential from the measured RFA IV-curves, i.e.\ pivot point, maximum of the derivative or linear fit, can be applied universally. With poor collimation the linear fit method is preferred. The most accurate result can be achieved by determining the pivot point from data collected with good collimation and then applying a correction for the systematic error, based on the prediction derived from the simulations, to de-convolute the effects caused by extraction optics and the RFA instrument.  
\end{itemize}

It should be noted, however, that while electrostatic focusing effects dominate the energy spreads after the extraction, the fraction of transverse to longitudinal momentum changes along the following beam line as the focusing ion optical elements, e.g. the dipole magnet, and the space charge forces change this ratio. However, the nonlinear electrostatic effects (presented for example in Fig.~\ref{fig:disc effect}) in the extraction generate also transverse emittance growth, which in practice is irreversible and makes the beam transport more challenging. Contrary to the electrostatic effect, the azimuthal magnetic force generated by the fringe field of the ECRIS solenoid always generates transverse emittance as demonstrated in Ref.~\cite{Wutte}. While the theory behind the magnetic emittance growth is well understood, the application in the ECRIS case is far from trivial as the extracted beam current and its distribution is expected to vary as a function of the extraction magnetic field. This has an effect on the dynamic plasma meniscus and associated electrostatic focusing or de-focusing as well as on the nonlinearities of the extraction system optics and the magnitude of the azimuthal magnetic kick. These effects can be large enough to explain the discrepancy between experiments and simple magnetic emittance theory reported in Ref.~\cite{Higurashi}.

Finally, it was discovered that the plasma potential of the CB ECRIS is affected by the microwave power and extraction field strength. Thus, it is concluded that these parameters affect the dynamics of the cold electron population, which is relevant for the formation of the ambipolar potential. The microwave power affects the energy content of the plasma, as shown in ref.~\cite{Noland} through diamagnetic loop measurements, whereas the extraction field is the most influential mirror point controlling the axial electron fluxes and losses~\cite{Izotov, Toivanen}. These two parameters can be expected to be the most influential in determining the charge breeding efficiency.

\section{Appendices}
\subsection{RFA resolution improvement} \label{RFADevelopment}

\begin{figure}[H]
    \centering
    \includegraphics[width=0.8\textwidth]{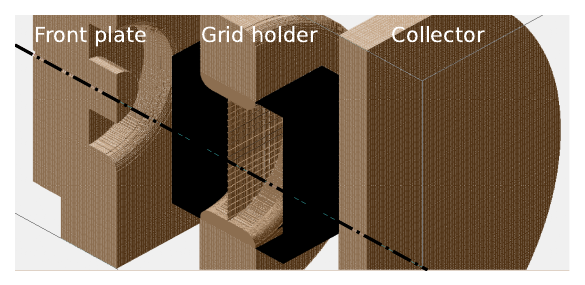}
    \caption{RFA \textsc{Simion} geometry cutview with the collimator on the left and the collector on the right, both at ground potential, and the grid and grid holder in the middle, at a variable potential. The black box shows the limits of the fine potential map.}
    \label{fig:SIMION geometry}
\end{figure}

The RFA was simulated with \textsc{Simion} using the geometry shown in figure~\ref{fig:SIMION geometry}. To simulate the effect of a fine meshing, we used two potential maps: a coarse geometry comprising the whole RFA and a fine geometry comprising only the grid, but with a much smaller pixel\footnote{\textsc{Simion}’s pixel are usually called grid units, but we won’t use this name while discussing an actual grid, so as to avoid confusion.} size. \textsc{Simion} uses the potentials calculated in the coarse map as boundary conditions on the fine map to calculate the potentials near the grid.

\begin{figure}[H]
    \centering
    \includegraphics[width=0.5\textwidth]{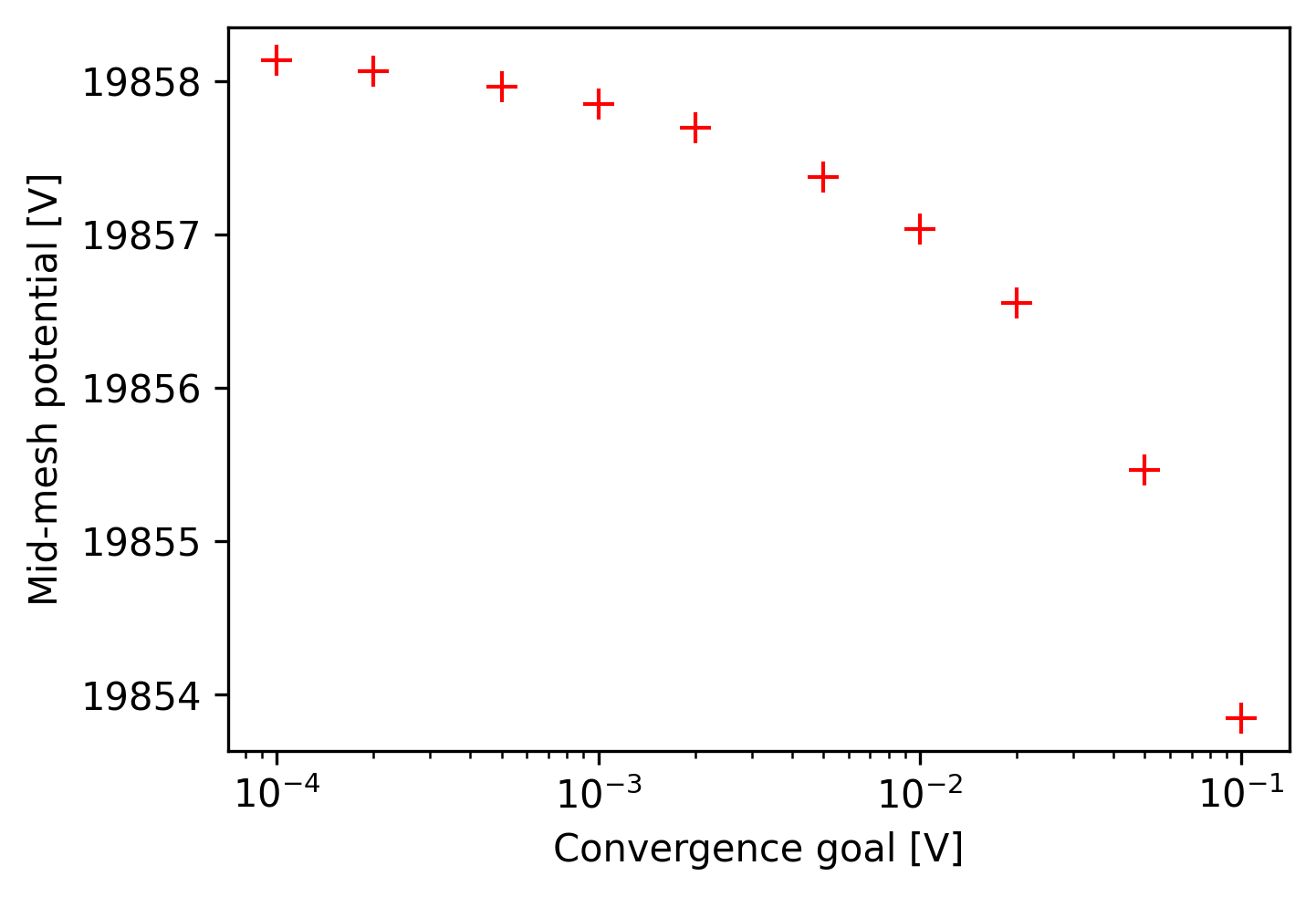}
    \caption{Potential at the center of a grid element as a function of the convergence goal of the Poisson solver in \textsc{Simion}.}
    \label{fig:grid refine goal}
\end{figure}

Nesting potential arrays is an effective way to reduce memory usage when only part of a geometry needs to be precisely described and/or does not share the same symmetries with the rest. Despite this, the calculation of the potential near the grid can still be time-consuming, especially when the convergence goal of the Poisson solver in \textsc{Simion} is very small. Figure~\ref{fig:grid refine goal} shows the calculated potential at the center of a grid element as a function of the convergence goal for a \SI{20}{\kilo\volt} potential applied to the grid. The first thing to notice is that the potential converges with lower convergence goals, and that this parameter only affects the potential near the grid by a few volts within the observed range. We chose a convergence goal of 0.001~V for all simulations, as a compromise between accuracy of the potential (around 1~V below the expected asymptotic value) and computation time (2--20 minutes per simulation depending on the thickness of the grid).

One can notice that the potential in the center of a grid element drops (which could be expected since the neighboring electrode are grounded), and that for the specific grid used for this calculation (1~mm grid step, 0.2~mm wire width) the drop is significant: around 140~V drop for 20~kV applied to the grid. This means that 20~keV ions could still cross the grid up to a grid voltage of 20140~V. This leads us to expect a large RFA distribution for this specific grid.

\begin{figure}[tbp]
    \centering
    \includegraphics[width=1.0\textwidth]{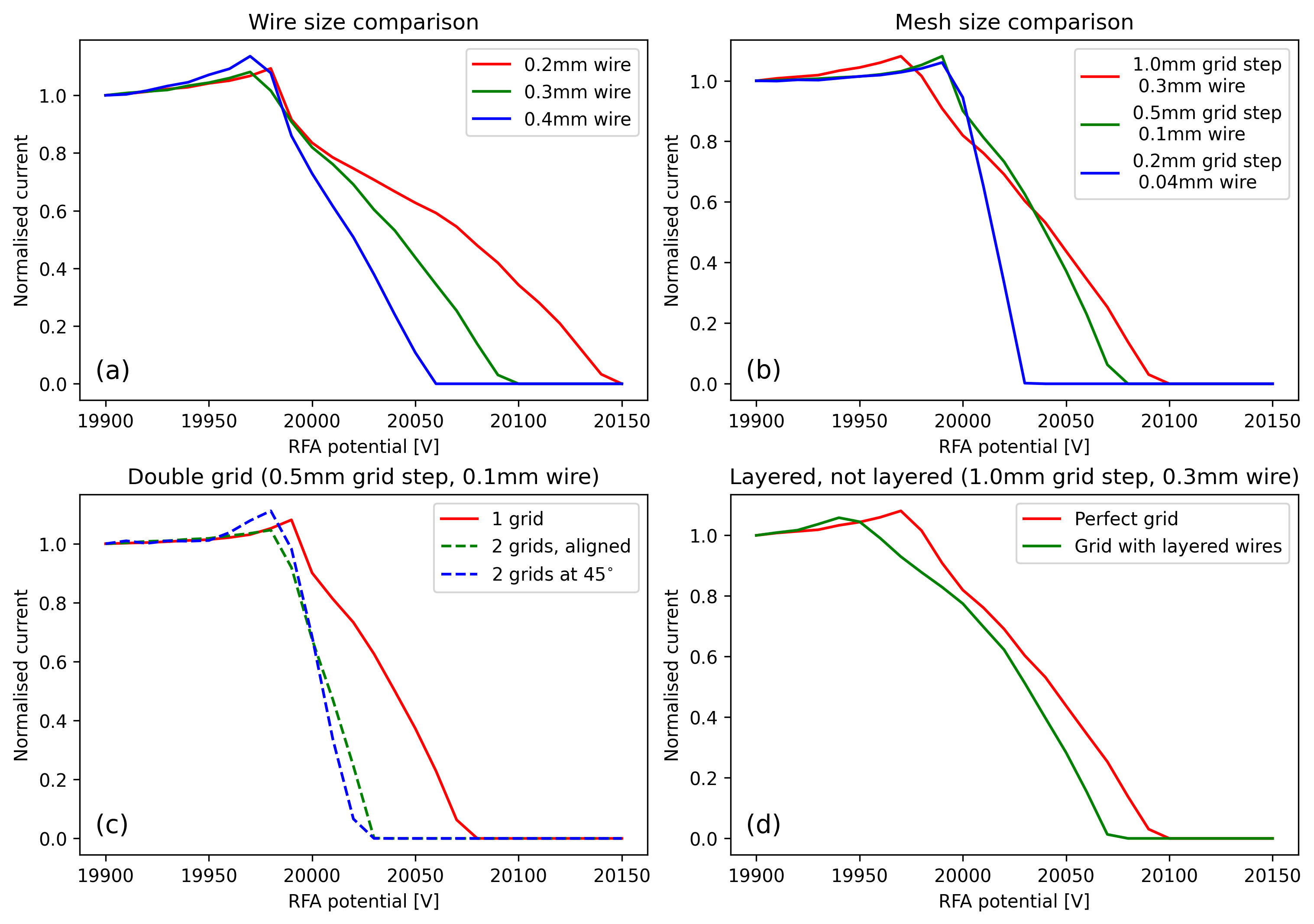}
    \caption{Normalised ion current vs grid potential for different grid parameters. (a) effect of the wire size, (b) effect of the mesh size, (c) effect of stacking two grids, and (d) difference between a "perfect grid", for which wires are merged on the same plane and a grid for which perpendicular wires are stacked on two different planes.}
    \label{fig:grid curves}
\end{figure}

The beam used as a reference is composed of 10000 $^{85}$Kr$^+$ ions at \SI{20}{\kilo\electronvolt} energy, created before the collimator, aligned with the beam axis with a distribution in position of \SI{1}{\milli\meter} RMS and \SI{0.2}{\degree} FWHM around the beam axis. With such distribution, about 1000 ions make it through the collimator, which is still a statistically significant sample. We simulated several grids in order to improve the RFA resolution. Figure~\ref{fig:grid curves} shows the ion current as a function of the grid potential for different grid configurations. All curves of figure~\ref{fig:grid curves} display the same general behaviour:

\begin{itemize}
    \item for $qV_S \ll E_0$ ($E_0$ being the initial kinetic energy) the ions cross the grid without deflection and the measured ion current depends very little on $V_S$.
    \item for $qV_S \lesssim E_0$ the ion current is above the expected maximum value. This is due to the voltage drop $V_d$ in the middle of a grid element, making each element a micro lens. This lens effect helps focusing the incoming ions into the grid holes, effectively increasing the grid transparency.
    \item for $qV_S > E_0 > q(V_S-V_d)$, ions flying near the center of a grid element still manage to cross the grid, but the ones close to the wires are reflected.
    \item for $q(V_S-V_d) > E_0$, no ion can fly through the grid.
\end{itemize}

Figure~\ref{fig:grid curves}(a) shows that the RFA resolution improves with increasing wire width, or more accurately with decreasing grid hole size, for a given grid step. Figure~\ref{fig:grid curves}(b) presents the effect of increasing the mesh density. As could be expected, a thinner mesh yields better RFA resolution because the voltage drop in a grid element is smaller. In figure~\ref{fig:grid curves}(c), we notice that stacking 2 grids improves the RFA resolution even further. In particular, two cross aligned grids (at 45°) result in a sharp decrease of the ion current and thus a better resolution. We noticed that for the mesh parameters of the real grid (\SI{1}{\milli\meter} wire spacing, \SI{0.3}{\milli\meter} wire diameter), the RFA resolution is worse in these simulations than in reality (see figure~\ref{fig:RFA spectrum Rb+Rb+ 092021}). To understand this discrepancy, we compared in figure~\ref{fig:grid curves}(d), simulation results for an ideal grid, for which perpendicular wires are on the same plane, and for a more realistic grid, made by stacking 2 planes of perpendicular wires. The resolution is indeed slightly better with this layered grid, but not enough to explain the discrepancy between simulation and reality. It should be noticed that the real grid is actually not layered but woven, which could have some effect on the resolution, but was difficult to simulate accurately. Nevertheless, in the end the goal of these simulations is not to perfectly reproduce reality but to help guiding small design changes to improve the RFA resolution.

\begin{figure}[tbp]
    \centering
    \includegraphics[width=0.5\textwidth]{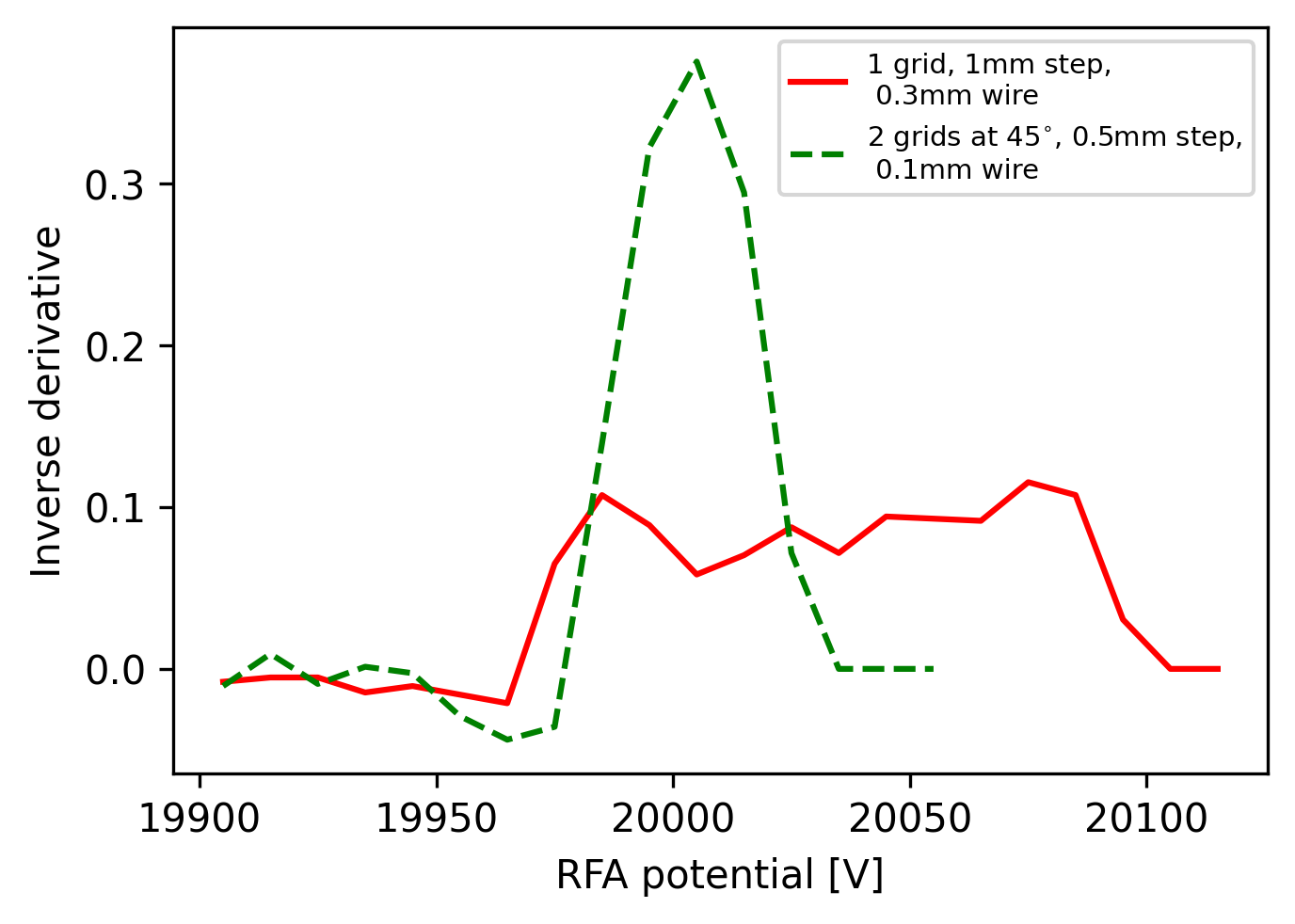}
    \caption{Derivate of the RFA current as a function of the grid potential for two grid configurations.}
    \label{fig:RFA resolution}
\end{figure}

Figure~\ref{fig:RFA resolution} shows the inverse derivative of the simulated RFA signal for the original grid and the best double grid configuration. Changing the original grid to a double grid with thinner meshing changes the FWHM resolution from around 120 V to 25 V, i.e.\ an improvement by a factor five.

Several other configurations have been simulated, including grids with a thinner mesh, double grid systems with space in between the grids or a thick plate with a dense array of holes instead of a grid, but will not be detailed here. In principle, it is possible to reach a better RFA resolution than in figure~\ref{fig:RFA resolution} by using still thinner meshes, or by spacing two grids by a few mm, although this would require some changes in the supporting structure.

\end{document}